\documentclass[usenatbib]{aa}  % print version
\usepackage[varg]{txfonts}

\usepackage[T1]{fontenc}
\usepackage{ae,aecompl}

\usepackage{graphicx}   % Including figure files
\usepackage{amsmath}    % Advanced maths commands
\usepackage{amssymb}    % Extra maths symbols
\usepackage{array}

\newcommand\jxllinewidth{8cm}
\newcommand\jxllinewidthb{8cm}
\newcommand\jxllinewidthD{14cm}

\def\trait{{\smallskip\hrule\smallskip}}

\def \ap#1{{#1}}

\def\trait{\noalign{\smallskip\hrule\smallskip}}
\def\norm#1{\left\Vert#1\right\Vert}
\def\abs#1{\left\vert#1\right\vert}
\def\norm#1{\Vert#1\Vert}
\def\Frac#1#2{{{\displaystyle\strut#1}\over{\displaystyle\strut#2}}}
\def\frac#1#2{{{#1}\over{#2}}}
\def\Dron#1#2{\Frac{\partial#1}{\partial#2}}
\def\dron#1#2{\frac{\partial#1}{\partial#2}}
\def\Der#1#2{\Frac{d#1}{d#2}}                                                

\def\m@th{\mathsurround=0pt}
\def\matrice#1{\left[\,\vcenter{\normalbaselines\m@th
    \ialign{\hfil$##$\hfil&&\quad\hfil$##$\hfil\crcr
    \mathstrut\crcr\noalign{\kern-\baselineskip}
    #1\crcr\mathstrut\crcr\noalign{\kern-\baselineskip}}}\,\right]}
\def\EQM#1{\vcenter{\normalbaselines\m@th
    \ialign{${\displaystyle ##}$\hfil&&\ ${\displaystyle ##}$\hfil\crcr
    \mathstrut\crcr\noalign{\kern-\baselineskip}
    \noalign{\smallskip}
    #1\crcr\mathstrut\crcr\noalign{\kern-\baselineskip}}}}

\def\crm{\cr\noalign{\medskip}}

\def\tu{{\bf\tilde u}}
\def\tr{{\bf\tilde r}}

\def\a{\alpha}
\def\bx{\bar x}

\def\l{\lambda}
\def\L{\Lambda}

\def\bea{\begin{eqnarray}}
\def\eea{\end{eqnarray}}
\def\be{\begin{equation}}
\def\ee{\end{equation}}

\def\llabel#1{\label{#1}} 
\def\cO{{\cal O}}
\def\Vec#1{\overrightarrow{#1}}

\def\G{{\bf G}}
\def\P{{\bf P}}

\def\cO{{\cal O}}
\def\dr{{\bf \dot r}}
\def\tu{{\bf\tilde u}}
\def\du{{\bf\dot u}}
\def\tr{{\bf\tilde r}}

\def\r{{\bf r}}                                                             
\def\u{{\bf u}}

\def\cO{{\cal O}}
\def\ga{\gamma}
\def\cC{{\mathscr{C}}}

\def\cD{{\mathcal{D}}}
\def\e{{\rm e}}
\def\tC{{\tilde{C}}}
\def\ve{\varepsilon}
\def\figdir{{./amdAAfigs}}
\def\figw{8 cm}
\def\phz{{\phantom{0}}}
\renewcommand\trait{\noalign{\smallskip\hrule\smallskip}}
\newcommand{\FFrac}[2]{{{\displaystyle\strut#1}\over{\displaystyle\strut#2}}}

\newcommand{\LEt}[1]{}

\def\nnb{\nonumber\\}
\def\av#1{\langle#1\rangle}
\def\H{\mathcal{H}}
\def\d{\mathrm{d}}
\def\Pr{\mathcal{P}}
\def\o{\mathrm{o}}
\def\O{\mathrm{O}}

\def\F{\mathcal{F}}
\newcommand{\Lim}[1]{\raisebox{0.5ex}{\scalebox{0.8}{$\displaystyle \lim_{#1}\;$}}}

\usepackage{color}
\usepackage{bm}%Bold math symbols
\usepackage{mathrsfs}

\def\totsta{70\ }
\def\totS{48\ }
\def\totW{22\ }
\def\totU{61\ }
\def\totnb{131\ }

\def\Sstatwo{42\ }
\def\Wstatawo{21\ }
\def\Ustatwo{34\ }
\def\tottwo{97\ }

\def\Sstathree{4\ }
\def\Wstathree{1\ }
\def\Ustathree{17\ }
\def\totthree{22\ }

\def\Sstafour{2\ }
\def\Wstafour{0\ }
\def\Ustafour{10\ }
\def\totfour{12\ }
\newcommand\tableA{
\begin{table*}[t]
        \begin{center}
                \caption{
                        Planetary distribution corresponding to different 
                        initial mass distribution. } 
                \label{table a_n}
\begin{tabular}{cccc}
        \trait
        $p$&   $a (n) $ & $m(a)$  & $m(n) $\\
        \trait
        $p\neq-3/2$ & $a^{\frac{2p+3}{6}} = a_0^{\frac{2p+3}{6}}  + \Frac{2p+3}{6}\left( \Frac{4\tC}{\zeta}\right)^{1/3} n$ & $\left(4\tC \zeta^2\right)^{1/3} a^{\frac{4p+3}{6}}$ & $m(n)\sim 4\tC \left(\Frac{4\tC}{\zeta}\right)^{\frac{-1}{2p+3}}\left(\Frac{2p+3}{6}n\right)^{2-\frac{3}{2p+3}}$ \\

        \trait
        $-\FFrac{3}{2}$  &$\log(a) =\log(a_0) +\left(\FFrac{4 \tC}{\zeta}\right)^{1/3} n $
        & $(4\tC\zeta^2)^{1/3}a^{-1/2}$& $ \log(m) \sim - \left(\Frac{\tC}{2 \zeta}\right ) ^{1/3} n$ \\
        \trait
        $0$ &$\sqrt{a}=\sqrt{a_0}+(4\tC/\zeta)^{1/3}\FFrac{n}{2}$ & $(4\tC\zeta^2)^{1/3}\,
        a^{1/2}$ & $ m \sim (2\tC^2 \zeta)^{1/3} n $\\
        \trait
\end{tabular} 
        \end{center}

\end{table*}  }

\newcommand\tableC{
\begin{table*}[t]
        \begin{center}
        \caption{Special values of $C_c(\a,\ga)$. The detail of the computations is provided in Appendix C.} 
        \llabel{formamd}
\begin{tabular}{ccccc}
\trait
$\ga$    &  $\a$    &  $e_c(\a,\ga) $  &  $e'_c(\a,\ga) $  &  $C_c((\a,\ga) $  \\
\trait
$\ga \rightarrow 0$   & $\a < 1/2 $  & $1-2\Frac{1-\a}{(1-2\a)^2} \ga^2$    & $1-2\a +2\Frac{\a(1-\a)}{(1-2\a)^2} \ga^2 $ &  $ 1-2\sqrt{\a(1-\a)} +\sqrt{\a}\ga  $ \\
$\ga \rightarrow 0$    &  $\a=1/2 $    &  $1- (4 \ga)^{2/3} $  &   $2^{1/3}\ga^{2/3}$     & $\Frac{\ga}{\sqrt{2}} $ \\
$\ga \rightarrow 0$    &  $\a>1/2 $    & $e_0   -\Frac{e_0}{\sqrt{\a(2\a-1)}}\ga $ & $\Frac{\sqrt{\a} e_0\ga }{\sqrt{2\a-1}}$ &    $\left(\sqrt{\a}-\sqrt{2-\frac{1}{\a}}\right)\ga$    \\    
    \trait
$\ga \rightarrow +\infty$& $0<\a <1 $ & $\Frac{1}{\ga}\Frac{1-\a}{\sqrt{2-\a}}$ & $1-\a-\Frac{1}{\ga}\Frac{\a (1-\a)}{\sqrt{2-\a}}$ & $1-\sqrt{\a(2-\a)} - \Frac{\sqrt{\a}(1-\a)^2}{2-\a} \Frac{1}{\ga}  
 $ \\
 \trait
$1$  & $0\leq\a \leq1 $   & $\Frac{1-\sqrt{1-\a+\a^2}}{\a}$ & $\sqrt{1-\a +\a^2} -\a $ & $1+\sqrt{\a} - \Frac{\sqrt{\a -2+ 2\sqrt{1-\a+\a^2}}}{\sqrt{\a}} $ \\
&&&& $-\sqrt{\a}\sqrt{1-2\a+2\sqrt{1-\a+\a^2}} $ \\
\trait
$\sqrt{\a}$ & $0\leq\a \leq1 $  & $ \Frac{1-\a}{1+\a} $  &  $ \Frac{1-\a}{1+\a} $  & $(1-\sqrt{\a})^2 $ \\
\trait
\end{tabular}

\end{center}
\end{table*}
}

\newcommand{\tableD}{
        \begin{table*}[t]
                
                \begin{center}
                        
                        \caption{Result of the analysis split in function of the multiplicity of the system}
                        \label{datarep}
                        \begin{tabular}{rrrrr}
                                \trait
                                Multiplicity & Strong stable & Weak stable & Unstable & Total \\ 
                                
                                \trait
                                2 & \Sstatwo & \Wstatawo & \Ustatwo & \tottwo \\ 
                                
                                3 & \Sstathree & \Wstathree & \Ustathree & \totthree \\ 
                                
                                4+ & \Sstafour & \Wstafour & \Ustafour & \totfour \\ 
                                \trait
                                Total & \totS & \totW & \totU & \totnb\\
                                \trait
                        \end{tabular}
                \end{center} 
\end{table*}}

\begin{document}

%\title{Why a super-Earth cannot be closer than 1200 AU ?}
\title{AMD-stability and the classification of planetary systems. }

\author{
 J. Laskar 
\and A. Petit 
}

\titlerunning{AMD-stability}
\authorrunning{Laskar \& Petit}

\institute{
 ASD/IMCCE, CNRS-UMR8028, Observatoire de Paris,  PSL, UPMC, 77 Avenue Denfert-Rochereau, 75014 Paris, France\\
\email{laskar@imcce.fr}
}

% These dates will be filled out by the publisher
\date{Accepted XXX. Received YYY; in original form ZZZ}

\abstract{
We present here in full detail the evolution of the angular momentum deficit (AMD) during collisions  
as it was described in \citep{Laskar2000}. Since then, the AMD has been revealed to be a key parameter  for the understanding of the 
outcome of planetary formation models. We define here  the AMD-stability criterion that can be easily 
verified on a newly  discovered planetary system. 
We show how AMD-stability can be used to establish a classification of  the  multiplanet systems in order 
to exhibit the planetary systems that are long-term stable because they are AMD-stable, 
 and those that are AMD-unstable which then require some additional dynamical studies to conclude on their stability. \LEt{ I would not use a hyphen in these cases: AMD stability, AMD stable, AMD unstable. If the acronym is spelled out you would not use a hyphen (angular momentum deficit stability) so I don't think one should be used in the short form. You have been consistent, however, so I have not changed them. Your choice// below: if exoplanet.eu is a URL it should go in a footnote.    } 
The AMD-stability classification is applied to the 131 multiplanet systems from The Extrasolar Planet Encyclopaedia  database (exoplanet.eu)
for which the orbital elements are sufficiently well known. 
}
%

% Select between one and six entries from the list of approved keywords.
% Don't make up new ones.

\maketitle

\section{Introduction}

The increasing number of planetary systems has made it necessary to search for a possible  classification of these 
planetary systems.  Ideally, such a classification  should  not require  heavy numerical  analysis
as it needs to be applied to large  sets of systems. 
Some possible approaches  can rely on the  stability analysis of these systems,  as this stability analysis is 
also part of the process used to consolidate the discovery of  planetary systems.
The stability analysis can also be considered  a key part to understanding  the wider question of the 
architecture of planetary systems.
In fact, 
the distances between planets and other orbital characteristic distributions is one of the oldest questions in 
celestial mechanics, the most famous attempts to set laws for this distribution of planetary
orbits being the   Titius-Bode power laws (see \citealt{Nieto1972,Graner1994} 
for a review).

 Recent research has focused on statistical analysis of observed architecture 
\citep{Fabrycky2012,Lissauer2011a,Mayor2011}, eccentricity distribution \citep{Moorhead2011,Shabram2015,Xie2016}, or 
inclination distribution \citep{Fang2012,Figueira2012}; see \cite{Winn2015} for a review.
The analysis of these observations  has been compared with models of system architecture 
\citep{Fang2013,Pu2015,Tremaine2015}.
These theoretical works usually use empirical criteria based on the Hill radius proposed by \cite{Gladman1993} and 
refined by \cite{Chambers1996,Smith2009}, and \cite{Pu2015}.
These criteria of stability  usually multiply the Hill radius by a numerical factor $\Delta_{\mathrm{sep}}$ 
empirically evaluated to a value around 10.
They are  extensions of the analytical results on Hill spheres for the three-body problem 
\citep{Marchal1982}. Works on chaotic motion caused by the overlap of mean motion resonances (MMR, 
\citealp{Wisdom1980,Deck2013,Ramos2015}) could justify the Hill-type criteria, 
but the results on the overlap of the MMR island are valid only for close orbits and for short-term stability.

Another approach to stability analysis is to consider the secular approximation of a planetary system. 
In this framework, the conservation of the semi-major axis leads to the conservation of another quantity, the angular momentum deficit (AMD; \citealp{Laskar1997,Laskar2000}).
An architecture model can be developed from this consideration \citep{Laskar2000}.
The AMD can be interpreted as a measure of the orbits' excitation \citep{Laskar1997} 
that limits the planets close encounters and ensures  long-term stability.\LEt{ sing-plural problem. I\ made it all singular. }
Therefore a stability criterion can be derived from the semi-major axis, the  masses and  the AMD of a system. 
In addition, it can be demonstrated that the AMD decreases during inelastic collisions (see section~\ref{sec.col}), accounting for the gain of stability of a lower multiplicity system.
Here we extend the previous analysis of \citep{Laskar2000}, and derive more precisely the AMD-stability criterion that can be used to establish a classification of the multiplanetary systems.

In the original letter \citep{Laskar2000},  the detailed computations were referred to as a preprint
to be published. Although this preprint has been in nearly final form for more than a decade, and has
even been provided to some researchers \citep{HernandezMena2011}, it is still unpublished. In Sects.~\ref{AMD}
and \ref{sec.AMD stab} we provide the fundamental concepts of AMD,  the full description, and all proofs for
the model that was described in \citep{Laskar2000}. \LEt{The assoc. editor suggested correcting this paragraph slightly. ok like so?}  
 This material is close to the unpublished preprint. 
Section~\ref{formation} recalls briefly the model of planetary accretion   based on AMD stochastic transfers  that was
first presented in \citep{Laskar2000}. This model  provides  analytical expressions 
for the averaged systems architecture and orbital parameter distribution, depending on the initial mass distribution
(Table \ref{table a_n}). 

In Sect.~\ref{classification},  we show how the AMD-stability criterion presented in section~\ref{sec.AMD stab} can be used to develop a classification of planetary systems. 
This  AMD-stability classification is then applied to a selection of   131 multiplanet systems from 
The Extrasolar Planet Encyclopaedia  database (exoplanet.eu)
\LEt{ in the Abstract you used lower case. Please be consistent }
with known eccentricities.
Finally, in Sect. 6 we discuss our results and provide some possible extension for this work.

\section{Angular momentum deficit  }
\label{AMD}
\subsection{Planetary Hamiltonian}

 Let $P_0, P_1, \dots , P_n$  be $n+1$ bodies of masses
 $m_0, m_1, \dots , m_n$ in gravitational interaction,  and let  $O$
be their centre  of mass. \LEt{ You show a slight preference for UK spelling conventions, and I have corrected accordingly.
 } For every body $P_i$, we  denote 
$\u_i = \Vec{OP_i}$. In the barycentric
reference frame with origin $O$, Newton's equations of motion form a
differential system of order  $6(n+1)$ and can be written in Hamiltonian form
using the canonical coordinates $(\u_i,\tu_i = m_i\dot \u_i)_{i=0,n}$ 
with Hamiltonian 

\be
H= \Frac{1}{2}\sum_{i=0}^{n}\Frac{\norm{\tu_i}^2}{m_i}
 -\mathcal{G}\sum_{0\leq i < j} \Frac{m_im_j}{\Delta_{ij}}
,\ee
where $\Delta_{ij}=\norm{\u_i-\u_j}$, and $\mathcal{G}$ is the constant of gravitation. 
The reduction of the centre of mass is achieved by using the canonical
heliocentric variables of Poincar\'e $( \r_i,\tr_i)$ \citep{Poin1905a,Laskar1995},
defined as  
\be
\r_0=\u_0 ; \qquad \r_i = \u_i-\u_0 \qquad\hbox{for } i\neq 0
\ee
\be
\tr_0 = \tu_0 +\tu_1+\cdots +\tu_n  ; \qquad \tr_i = \tu_i \qquad\hbox{for
} i\neq 0 
\ee

This Hamiltonian can then be split into an integrable part, $H_0$, and a perturbation, $H_1$, 

\be
H = H_0 +H_1 
\ee
with
\be
H_0 =  \Frac{1}{2}\sum_{i=1}^{n}\Frac{\norm{\tr_i}^2}{m_i}
-\mathcal{G}\sum_{i=1}^n\Frac{m_0m_i}{r_i}
\ee
and 
\be
H_1 =\Frac{1}{2} \Frac{\norm{\tu_0}^2}{m_0} -\mathcal{G}\sum_{1\leq i < j} \Frac{m_im_j}{\Delta_{ij}}
\ee

The integrable part, $H_0$, \LEt{ ok? to avoid the symbol at the beginning of the sentence} is the Hamiltonian of a sum of disjoint Kepler problems  of a single
planet of mass $m_i$ around a fixed Sun of mass $m_0$. A
set of adapted  variables for $H_0$ will thus be given by the 
elliptical elements,  $(a_k,e_k,i_k,\lambda_k, \varpi_k, \Omega_k)$, where
$a_k$ is the semi-major axis, $e_k$  the eccentricity, $i_k$ the inclination,  
$\lambda_k$ the mean longitude, $\varpi_k$ the longitude of
the perihelion, and $\Omega_k$ the longitude of the node.
They are defined as the elliptical elements associated to the Hamiltonian 
\be
H_{0k} = \Frac{1}{2} \Frac{\norm{\tr_k}^2}{m_k}
-\mu\Frac{m_k}{r_k}
\ee
with $\mu = \mathcal{G}m_0$. 

\subsection{Angular momentum deficit (AMD)}
The total linear momentum of the system is null in the barycentric reference frame
\be
L = \sum_{i=0}^n m_i\dot\u_i   = \sum_{i=0}^n \tu_i = \tr_0= 0 \ .
\ee

Let $\G$ be the total  angular momentum. Its expression is not changed 
by the linear symplectic change of variable $(\u,\tu)\longrightarrow (\r,\tr)$. We have thus
\be
\G = \sum_{i=0}^n \u_i\,\wedge \tu_i  = \sum_{i=1}^n \r_i\,\wedge \tr_i
\ee
When expressed in heliocentric variables, the angular momentum is thus the sum 
of the angular momentum of the Keplerian problems of the unperturbed 
Hamiltonian $H_0$.
In particular, if the angular momentum direction is assumed to be the axis $z$, the norm of the angular momentum 
is
\be
G = \sum_{k=1}^n \L_k\sqrt{1-e_k^2}\cos i_k
,\ee
where $\L_k =m_k\sqrt{\mu\, a_k}$. For such a system, the  AMD 
is defined as the difference between 
the norm of the angular momentum of a coplanar and circular system with 
the same semi-major axis values and 
the norm of the angular momentum ($G$),  i.e. (Laskar, 1997, 2000)
\be
C =  \sum_{k=1}^n \L_k\left(1-\sqrt{1-e_k^2}\cos i_k\right)
\ee
\subsection{AMD and collision of orbits}
\label{sec.col}

The instabilities of a planetary system often result in a modification of its architecture.
A planet can be ejected from the system or can fall into the star;  in both cases this results in a 
loss of  AMD for the system.
The outcome of the AMD after a planetary collisions is less trivial and  needs to be computed.
Assume that among our $n+1$ bodies, the (totally inelastic) collision of  two bodies of 
masses $m_1$ and $m_2$, and orbits  $\cO_1,\cO_2$  occurs, forming  a new body
$(m_3,\cO_3)$. During this collision we assume that  the other bodies are not affected.
The mass is conserved 
\be
m_3 = m_1+m_2,
\llabel{eq.12}
\ee
and the linear momentum in the barycentric reference frame is conserved 
so $\tu_3 =  \tu_1 +  \tu_2 $, and also 
\be
\tr_3 =  \tr_1 +  \tr_2 \ ;
\llabel{eq.13}
\ee
On the other hand,   $\r_3 = \r_1 = \r_2$ at the time of the collision, so the
angular momentum is also conserved
\be
 \r_3\,\wedge \tr_3 = \r_1\,\wedge \tr_1 + \r_2\,\wedge \tr_2 
\llabel{eq.ang.1}
\ee

It should be noted that the transformation of the orbits $(m_1,\cO_1) + (m_2,\cO_2)
\longrightarrow (m_3,\cO_3)$ during the collision is perfectly defined by
Eqs. (\ref{eq.12},~\ref{eq.13}). The problem which remains is to compute the
evolution of the elliptical elements during the collision.

\subsubsection{Energy evolution during  collision}
Just before  the collision, we assume that the orbits  $(m_1,\cO_1)$ and $(m_2,\cO_2)$
are elliptical heliocentric orbits. At the time of the collision, only these two bodies
are involved and the other bodies are not affected. The evolution of the orbits are
thus given by the conservation laws (\ref{eq.12},~\ref{eq.13}). The Keplerian energy of
each particle is 
\be
h_i = \Frac{1}{2} \Frac{\norm{\tr_i}^2}{m_i} -\mu\Frac{m_i}{r_i}=-\mu\Frac{m_i}{2a_i}
\ee
At collision, we have $\r_1=\r_2=\r_3=\r$; we have thus the conservation of the
potential energy
\be
-\mu\Frac{m_3}{r_3}= -\mu\Frac{m_1+m_2}{r}=-\mu\Frac{m_1}{r_1}-\mu\Frac{m_2}{r_2} \ .
\ee
The change of Keplerian energy is thus given by the change of kinetic energy
\be
\delta h = h_3-h_1-h_2 =  \Frac{1}{2} \Frac{\norm{\tr_3}^2}{m_3}
-\Frac{1}{2} \Frac{\norm{\tr_1}^2}{m_1}-\Frac{1}{2} \Frac{\norm{\tr_2}^2}{m_2}
;\ee
that is, with Eqs. (\ref{eq.12},~\ref{eq.13}),  
\be
\EQM{
&2m_1m_2m_3\delta h  \crm
&\phz = m_1m_2(\tr_1 + \tr_2)^2-m_2(m_1+m_2)\tr_1^2 - m_1(m_1+m_2)\tr_2^2 \crm
&\phz = -m_2^2\tr_1^2 -m_1^2\tr_2^2 +2m_1m_2\tr_1\cdot \tr_2 \crm
&\phz =-(m_2\tr_1-m_1\tr_2)^2 \crm
&\phz \leq 0 
}
\ee
Thus, the Keplerian energy of the system decreases during collision. Part of the kinetic
energy is dispelled  during the collision. As expected, there is no loss of energy when
$m_2\tr_1=m_1\tr_2$, that is, as $m_1m_2\neq 0$, when $\dot \u_1=\dot \u_2$. As an immediate consequence of the decrease of energy during the
collision, we have
\be
\Frac{1}{a_3} \geq \Frac{m}{a_1} +\Frac{1-m}{a_2}
\llabel{eq.15}
\ee
with 
\be
m=\Frac{m_1}{m_1+m_2} = \Frac{m_1}{m_3}
\ee

\subsubsection{AMD evolution during  collision}
Let $f(x) = 1/\sqrt{x}$. As $f'(x) <0 $ and $ f''(x)>0$,  we have, as 
$f$ is decreasing and convex,

\be
f\left(\Frac{1}{a_3}\right) \leq f\left( \Frac{m}{a_1} +\Frac{1-m}{a_2}\right)
\leq   m f\left( \Frac{1}{a_1}\right)  +(1-m)f\left(\Frac{1}{a_2}\right)
\label{eq.21}
\ee
thus

\be
m_3 \sqrt{a_3} \leq m_1 \sqrt{a_1}  + m_2 \sqrt{a_2}
\ee
During the collision, the angular momentum is conserved (\ref{eq.ang.1}), and so is
 the conservation of its normal component, that is 
\be
\EQM{
m_3 \sqrt{\mu a_3}\sqrt{1-e_3^2}\cos i_3 =  \crm
m_1 \sqrt{\mu a_1}\sqrt{1-e_1^2}\cos i_1 + m_2 \sqrt{\mu a_2}\sqrt{1-e_2^2}\cos i_2 \ .
}
\llabel{eq.angz.1}
\ee
We deduce that in all circumstances we have a decrease of the  angular momentum deficit
during the collision, that is 
\be
C_3 \leq C_1+ C_2 
\llabel{eq.amd.1}
\ee
with
\be
C_k =  m_k \sqrt{\mu a_k}(1-\sqrt{1-e_k^2}\cos i_k ) \qquad {(k=1,3)}
\ee

The equality can hold in (\ref{eq.amd.1}) only if $m_1=0$,
$m_2=0$, or $a_1=a_2$ and $\du_1=\du_2$, that is when one of the bodies is 
massless, or when the two bodies are on the same orbit and at the same position
(at the time of the collision, we also have  $\r_1=\r_2$). 

The diminution of AMD during collisions acts as a stabilisation of the system.
A parallel can be made with thermodynamics, the AMD behaving for the orbits like the kinetic energy for the molecules of a perfect gas.
The loss \ap{of} AMD during collisions can thus be interpreted as a cooling of the system.

\section{AMD-stability} 
\llabel{sec.AMD stab}
We   say that a planetary system is AMD-stable if the angular momentum deficit (AMD)
amount in the system is not sufficient  to allow for planetary collisions. 
As this quantity is conserved in the secular system at all orders (see Appendix~\ref{annex-average}), \LEt{ this should be changed throughout. Also in Table 1 } we 
conjecture  that in
absence of short period  resonances, the AMD-stability ensures the practical\footnote{
Here practical stability means  stability over a very long time compared to the expected life of 
the central star.} long-term
stability of the system. 
\LEt{ I think long-term/short-term stability is the appropriate expression. 
If you keep long time/short time you must use the hyphen consistently. All with or all without.   } 
Thus for an AMD-stable system, short-term stability will imply
long-term stability.

The condition of AMD-stability is obtained when the orbits of two  planets 
of semi-major axis $a, a'$ cannot intersect  under the assumption  
that the total AMD $C$  has been absorbed by 
these two planets. It can be seen easily that the limit
condition of collision is obtained in  the planar case and can thus  be
written as 
\begin{eqnarray}
a(1+e) = a'(1-e')   
\llabel{inter1} \\ 
m \sqrt{\mu a} (1-\sqrt{1-e^2}) + m' \sqrt{\mu a'} (1-\sqrt{1-e'^2}) = C 
\llabel{inter2}
,\end{eqnarray}
where $(m,a,e)$ are the parameters of the inner orbit and $(m',a',e')$ those of the outer orbit ($a\leq a'$).

\subsection{Collisional condition}
We assume that $a,a', m, m'$ are non-zero. Denoting $\a = a/a'$, $\ga = m/m'$, 
the system in Eq. (\ref{inter2}) becomes
\bea
\cD(e,e') &=\a e +e' - 1+\a =0  
\llabel{droite}
\eea
\bea
\EQM{
\cC(e,e')&=\gamma \sqrt{\a} (1-\sqrt{1-e^2}) +  (1-\sqrt{1-e'^2}) \crm
&= C /\Lambda' 
\llabel{amd}
}
\eea
with $\Lambda' = m'\sqrt{\mu a'}$, and where  
 $\cC(e,e')=C /\Lambda' $ is called the relative AMD.
As $e$ and $e'$ are 
planetary eccentricities, we also have   
\be
0\leq e \leq 1 \ ;\qquad  0\leq e' \leq 1 \ .
\llabel{eccen}
\ee

\begin{figure}[h]
\includegraphics*[width=\figw]{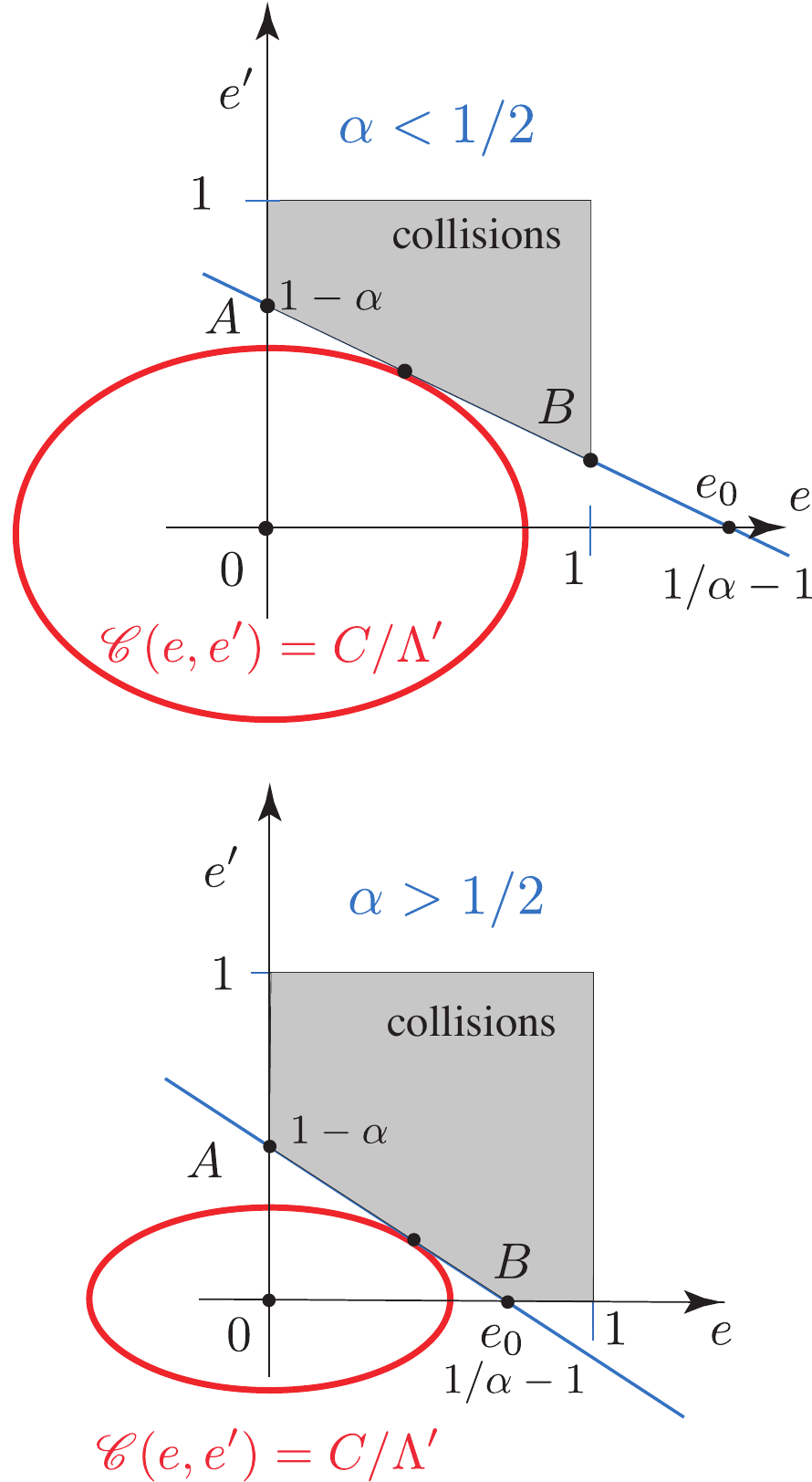}  
\caption{Collision conditions for $e_0 = 1/\a-1$. Case (a): $\a< 1/2 \Longleftrightarrow  e_0 > 1$.
Case (b): $\a> 1/2 \Longleftrightarrow  e_0 < 1$. Collisions can only occur  in the shaded
region.}
\label{amd1.fig1} 
\end{figure}

The set of collisional conditions (Eqs.~\ref{droite},~\ref{amd},~\ref{eccen}) 
can be solved using Lagrange multipliers. We are looking for the minimum value of 
the relative  AMD $\cC(e,e')$ (\ref{amd}) for which the collision condition (\ref{droite}) is satisfied.  These conditions are visualised in the $(e,e')$ plane 
in Fig.~\ref{amd1.fig1}. The collision condition (\ref{droite}) corresponds to  the segment $AB$ 
of Fig.~\ref{amd1.fig1}. The domain of collisions is  limited by the conditions
(\ref{eccen}). For $e = 0$, we have $e'_0 = 1-\a < 1$, and  
the intersection  of the collision line with the axis $e'=0$ is obtained for 
$e_0 = {1}/{\a}-1 $.
This value can be 
greater or smaller than $1$ depending on the value of $\a$. 
We have thus  the different  cases (Fig.~\ref{amd1.fig1})
\bea
(a)\quad \a &< \Frac{1}{2} \Longleftrightarrow  e_0 > 1 \quad &\hbox{and} \quad   e \leq 1 \\
(b) \quad \a &> \Frac{1}{2} \Longleftrightarrow  e_0 < 1 \quad &\hbox{and} \quad   e < e_0  
\eea
and the limit case $\a = 1/2$, for which $e_0 =1$.
In all cases, the  Lagrange multipliers condition is written 
\be
\lambda \nabla  \cD(e,e')  =  \nabla \cC(e,e') 
,\ee
which gives 
\be
\Frac{\sqrt{1-e'^2}}{e'} = \Frac{\sqrt{\a}}{\ga}\Frac{\sqrt{1-e^2}}{e} \llabel{eprim}
\ee

This relation allows  $e'$ to be eliminated in the collision condition (\ref{droite}), which
becomes an equation in the single variable $e$, and parameters $(\a,\ga)$:
\be
F(e,\a,\ga) = \a e + \Frac{\ga e}{\sqrt{\a(1-e^2) +\ga^2e^2}} -1 +\a =0
\llabel{colli}
.\ee
Here $F(e,a,\ga)$ is properly defined for $(e,a,\ga)$ in the domain $D_{e,\a,\ga}$ defined by 
 $e\in [0,1], \a \in ]0,1], \ga\in]0,+\infty[$, as in this
domain, $1-e^2 +\ga^2e^2/\a >0$. We also have 
\be
\Dron{F}{e}(e,\a,\ga) = \a + \Frac{\a\ga}{\left(\a(1-e^2) +\ga^2e^2\right)^{3/2}}  \ .
\ee
Thus, $\Dron{F}{e}(e,\a,\ga) > 0$ on the domain  $D_{e,\a,\ga}$ and $F(e,\a,\ga)$ is strictly
increasing with respect to $e$ for 
$e \in [0,1]$.
Moreover, as
$0<\a<1$, 
\be
F(0,\a,\ga) = -1+\a < 0 \ ;\qquad F(1,\a,\ga) = 2\a >0 \ ;
\ee
and 
\be
F(e_0,\a,\ga) = \Frac{\ga e_0}{\sqrt{\a\left(1-e_0^2\right) +\ga^2e_0^2}} > 0 \ .
\ee
The equation of collision (\ref{colli})  thus always has a 
single solution $e_c$ in the interval $]0,\min (1,e_0)[$. This ensures that this 
critical value of $e$ will  fulfil the condition (\ref{eccen}).
The corresponding value  of the relative AMD $C_c (\a,\ga)= \cC(e_c,e'_c)$ is then obtained 
through (\ref{amd}).

\subsection{Critical AMD $C_c (\a,\ga)$}
\label{C_c}
We have thus demonstrated that for a given pair of ratios of semi-major axes and 
masses,  $(\a,\ga)$, there is always a unique  
critical value   $C_c (\a,\ga)$ of the relative AMD  $\cC=C /\Lambda' $ 
which defines the AMD-stability. The system of two planets is AMD-stable if and only if 
\be
\cC=\Frac{C }{\Lambda'} <  C_c (\a,\ga) \ .
\llabel{critic}
\ee
 
The value of the critical AMD  $C_c (\a,\ga)$ is obtained by computing first  the critical 
eccentricity $e_c (\a,\ga)$ which is the unique solution of the collisional equation (\ref{colli})
in the interval $[0,1]$. The critical AMD is then $C_c (\a,\ga)=\cC(e_c,e'_c)$ (Eq.\ref{amd}), where
the critical value  $e'_c$ is obtained  from $e_c$ through  Eq. \ref{droite}.    
It is important to note that the critical AMD, and thus the AMD-stability condition, 
 depends only on $(\a,\ga)$.

\subsection{Behaviour of the critical AMD}

\tableC

\begin{figure}
        \begin{center}
\includegraphics*[width=\figw]{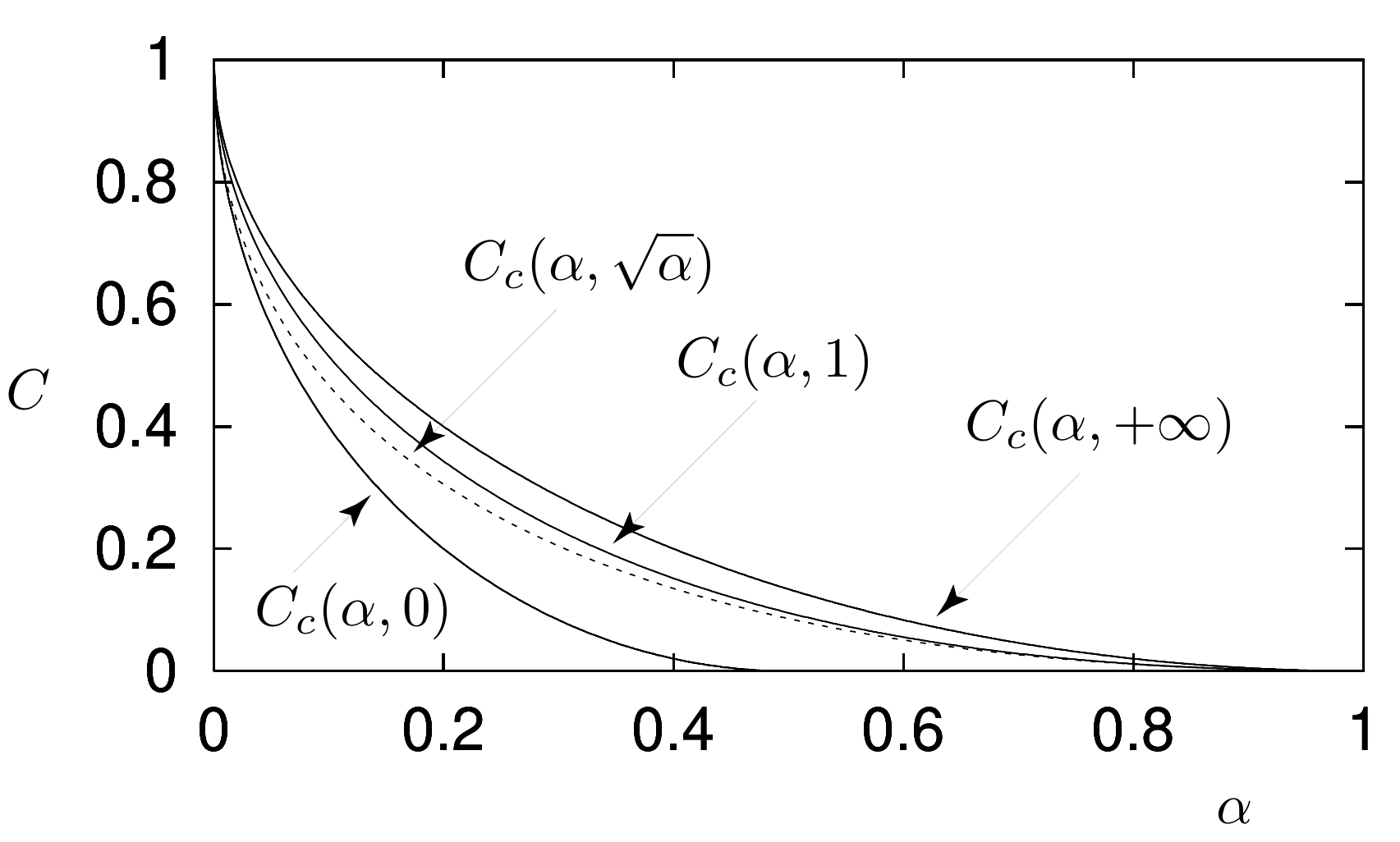}  
\caption{Values of  $C_c (\a,\ga)$ vs. $\a$ for the different $\ga$ values 
for which an explicit expression of $C_c (\a,\ga)$  is obtained.}
\label{amd1.fig2}       
\end{center}
\end{figure}

We  now analyse the general properties of the critical AMD function $C_c (\a,\ga)$.
 As $\dron{F}{e}(e,\a,\ga) > 0$, we can  apply
the implicit
function theorem to the domain  $D_{e,\a,\ga}$,  which then ensures that the solution of the collision equation (\ref{colli}), $e_c(\a,\ga)$, is a continuous function of
$\ga$ (and even analytic for $\ga \in ]0,+\infty[$). Moreover, on  $D_{e,\a,\ga}$,
\be
\Dron{F}{\ga}(e,\a,\ga) = \Frac{\a e\left(1-e^2\right)}{\left(\a\left(1-e^2\right) +\ga^2e^2\right)^{3/2}} \geq 0 \ .
\ee
We also have 
\be
\Dron{e_c}{\ga}(\a,\ga) = - \Frac{e_c\left(1-e_c^2\right)}{\left(\a\left(1-e_c^2\right)+\ga^2e_c^2\right)^{3/2}+\ga} \leq 0
\llabel{amds1.eq1}
,\ee 
and $e_c(\a,\ga)$ is a decreasing function of $\ga$.
For any given values of the semi-major axes ratio $\a$, and masses $\ga$, we can thus find 
the critical value $C_c (\a,\ga)$ which allows for a collision (\ref{critic}). For the
critical value  $C_c (\a,\ga)$, a single solution  corresponds to the
tangency condition (Fig.~\ref{amd1.fig1}), and this solution is obtained at the critical value $e_c(\a,\ga)$ for
the eccentricity of the orbit $\cO$.
The values of the critical relative AMD  $C_c (\a,\ga)$ 
are  plotted in Figure~\ref{amd1.fig2} versus $\a$, for different values of $\ga$. 
Deriving Eq.~\ref{amd} with respect to $\ga$, one obtains 
\be
\EQM{
\Dron{\cC}{\ga} = &\sqrt{\a}\left(1-\sqrt{1-e^2}\right) \crm
&+ \ga\sqrt{\a}\Frac{e}{\sqrt{1-e^2}}\Dron{e}{\ga}
+ \Frac{e'}{\sqrt{1-e'^2}}\Dron{e'}{\ga} \ .
}
\ee

Using the two relations (\ref{droite},~\ref{eprim}), this reduces to 
\be
\Dron{C_c(\a,\ga)}{\ga} = \sqrt{\a}\left(1-\sqrt{1-e_c^2}\right) > 0
\llabel{eq.dcga}
\ee
Thus $C_c(\a,\ga)$  strictly increases with $\ga$.
In the same way 
\be
\Dron{C_c(\a,\ga)}{\a} = \Frac{\ga}{2\sqrt{\a}} \left( 1- \Frac{\left(1+e_c^2\right)}{\sqrt{1-e_c^2}}\right)
<0
\llabel{eq.dca}
\ee
and  $C_c(\a,\ga)$   decreases with $\a$.

Now that the general behaviour of the critical AMD $C_c(\a,\ga)$  is known, we can specify its 
explicit expression in a few  special cases that are displayed in Table~\ref{formamd}.
The computations and the higher order developments can be found in Appendix~\ref{Annex C_c}. 

A development of $C_c(\a,\ga)$ for $\a \longrightarrow 1$ can also be obtained (see Appendix~\ref{Annex C_c}). With  $\eta = 1-\a$, we have
\be
C_c(\a,\ga)  \sim \Frac{\gamma}{\gamma+1} \Frac{ \eta^2 }{2} +\O(\eta^3) \ .
\ee

We  now present two applications of the AMD-stability, the formation of planetary systems, and a classification of observed planetary systems.

\section{AMD model of planet formation}
\label{formation}

\ap{Once the disc \LEt{  disc  (UK) / disk (US), except for computer disks which is always spelled with ``k".   \     } has been depleted, the last phase of terrestrial planet formation begins with a disc
composed of planetary embryos and planetesimals \citep{Morbidelli2012}.
In numerical simulations of this phase, the AMD has been used as a statistical measure for comparison with the inner solar system \citep{Chambers2001}.
In this part, we recall the very simple model of embryo accretion described in \cite{Laskar2000} interpreting the $N$-body dynamics as AMD-exchange.}

\subsection{Collisional evolution}
So far we have not made special simplifications, and the model simply preserves the mass and 
the momentum in the barycentric reference frame.
We  make   an additional assumption  here: between collisions, 
the evolution of the orbits is similar to the evolution they would have in 
the averaged system in the presence of chaotic behaviour. The  orbital parameters will thus evolve 
in a limited manner, with a stochastic diffusion process, bounded by the conservation of the  total AMD.
As shown in section~\ref{sec.col}, during a collision  the AMD  decreases (eq.~\ref{eq.amd.1}). 
\ap{We assume the collisions to be totally inelastic with perfect merging. }
Indeed, \cite{Kokubo2010} \ap{and \cite{Chambers2013}} have shown  that the detailed mechanisms of collisions, \ap{such as the possibilities of hit-and-run  or fragmentation of embryos, }barely 
change the final architecture of simulated systems.\LEt{ have I\ interpreted correctly here? I changed singular/plural and a comma }
The total AMD of the system will thus be constant between collisions, and will decrease during collisions. 
On the other hand, the AMD for a particle is of the order of 
$\frac{1}{2}m\sqrt{a}e^2$. As the mass of the particle increases, its excursion
in eccentricity will be more limited, and fewer collisions will be possible. 
The collisions will stop when the total  AMD of the system is too small 
to allow for planetary collisions.

In the accretion process, we consider a planetesimal of semi-major axis $a$ and its immediate
neighbour, defined as the planetesimal with semi-major axis $a'$, such that 
there is no other planetesimal with semi-major axis between $a$ and $a'$.
In this case we can assume that $\a$ is 
 close to $1$ and, as explained in Appendix~\ref{Annex C_c}, we  use  an approximation of the
critical AMD value
$C_c(\a,\gamma)$:
\be
C_{c1}(\a,\ga)=k(\ga) \left(\Frac{\delta_a}{a}\right)^2
\llabel{eq.cequiv2}
,\ee
where $\delta_a=a'-a$ and  $k(\ga)=\ga/(2(\ga+1)) $ .

\subsection{AMD-stable planetary distribution}

In this section, we search for the laws followed by the planetary distribution 
of a model formed following the above assumptions. We thus start from an arbitrary 
distribution of mass of planetesimals $\rho(a)$, and let the system evolve 
under the previous rules.
We search for the condition of AMD-stable planetary systems, 
obtained by random accretion of planetesimals. This condition requires that
the final AMD value cannot allow for planetary crossing among the planets.
Let  $C$ be the value of the AMD at the end of the accretion process.
If we consider the planetesimal of semi-major axis
$a$, its mass will continue to increase by accretion with a body of semi-major axis $a' > a$, as long as
\be
\cC=\Frac{C}{\Lambda'} \geq C_{c1} = k \left(\Frac{\delta_a}{a}\right)^2
\llabel{eq.sca1}
\ee

The initial linear density of mass is $\rho(a)$. As $a'$ is the closest neighbour to $a$, we can
assume that all the planetesimals initially between $a$ and $a'$ have been absorbed by the two 
bodies of mass $m(a)$ and $m(a')$. 
At first order with respect to  $\delta_a / a$, we have  thus 
\be
  m(a') \sim m(a) \sim  \rho(a) \delta_a 
\llabel{eq.sca2}
,\ee
and from (\ref{eq.sca1})  at the limiting case we have 
\be
\Frac{\tC}{\delta_a \rho(a) \sqrt{a}} = k  \left(\Frac{\delta_a}{a}\right)^2
,\ee 
where $\tC= C/\sqrt{\mu}$. We have thus 
\be
\delta_a = \left(\Frac{\tC}{k}\right)^{1/3} a^{1/2}\rho^{-1/3}
\llabel{eq.scaa}
,\ee
and from (\ref{eq.sca2}) 
\be
m(a) = \left(\Frac{\tC}{k}\right)^{1/3} a^{1/2}\rho^{2/3}
\llabel{eq.scam}
.\ee

\subsection{Scaling laws with initial mass distribution $\rho(a) \propto a^p$.} 

Using the previous relations, we can compute the resulting systems for various initial 
mass distribution, in particular for $\rho(a) = \zeta a^p$. 
From equation (\ref{eq.scam}), we obtain for two consecutive planets 
\be
\Frac{m}{m'} = \a^{1/2} \left(\Frac{\rho(a)}{\rho(a')}\right)^{2/3} = \a^{\frac{1}{2}+\frac{2}{3}p}
\ee
and from Eq. (\ref{eq.cequiv2}), as $\lim_{\a \longrightarrow 1} \ga =1$, we thus 
have $k(\ga) =  \Frac{1}{4} $ in all cases.
Relation (\ref{eq.scaa}) can be rewritten 
\be
\delta_aa^{\frac{p}{3}-\frac{1}{2}} = (4\tC)^{1/3}\zeta^{-1/3}\delta n\, ,
\label{eq.sca3}
\ee
where $\delta n = 1$ is the increment from planet $a$ to $a'$. By integration, this 
difference relation becomes for $p\neq -\frac{3}{2}$,

\be
a^{\frac{2p+3}{6}} = a_0^{\frac{2p+3}{6}}  + \frac{2p+3}{6}\left( \Frac{4\tC}{\zeta}\right)^{1/3} n \ .
\ee
For $p=-\frac{3}{2}$, we obtain 
\be
\log(a) =\log(a_0) +(4\tC)^{1/3} \zeta^{-1/3} n.
\ee

In particular, for $p=0$ (constant distribution and $\rho(a) = \zeta$), from (\ref{eq.sca3}), we  have  
\be
\sqrt{a} = \sqrt{a_0} + \left( \Frac{\tC}{2\zeta}\right)^{1/3} n \ .
\ee

For the masses,  from (\ref{eq.scam}), we have  for large $n$ (or if $a_0$ is small )
\be
m(n) \sim (2\tC^2\zeta)^{1/3} n \ .
\ee

For $p=-3/2$,  we find a power law similar to the  Titius-Bode law for  planetary distribution\footnote{\ap{
The value  $p=-3/2$ corresponds to a surface density proportional to $r^{-5/2}$, which is different from 
 the $-3/2$ surface density exponent  of the minimum mass solar nebula \citep{Weidenschilling1977}. 
For the solar system, \citep{Laskar2000} found $p=0$ to be the best fitting value; 
it corresponds to  a surface density proportional to $r^{-1}$.}}.
The different expressions deduced from this model of planetary accretion are listed 
\LEt{ i.e. added together? or are summarised, are given, are listed  } in Table~\ref{table a_n}.
\def\trou{\noalign{\medskip}}

\tableA

The previous analytical results were tested on a numerical model of our accretion scheme \citep{Laskar2000}.
The model was designed to fulfil the conditions (\ref{eq.12},\ref{eq.13}) of section~\ref{sec.col}.
Five thousand  simulations  were started with a large number of orbits (10 000) and followed in order to  look for orbit intersections. 
When an intersection occurs,  the two bodies merge into a new one whose orbital parameters are determined 
by the collisional equations (\ref{eq.12},\ref{eq.13})  (see Appendix A). Between collisions, the orbits do not
evolve, apart from a diffusion of their  eccentricities, which fulfils the condition of conservation
of the total AMD. This is roughly what would occur  in a chaotic   secular motion.

The main parameter of these simulations is the final AMD value, $C$. Because \LEt{ While? } the AMD decreases during
collisions, 
and in order to obtain final systems with a given value $C$ of the AMD,
the eccentricities were increased by a small amount in order to raise the AMD to the desired final value.
This is justified as close encounters can also increase the AMD value.
Indeed, \ap{$N$-body simulations \citep{Chambers2001,Raymond2006}
present a first phase of AMD increase at the beginning of the simulations 
before the AMD  decreases and converges to its final value.}
These simulations were extremely rapid to integrate as
the orbital motion of the orbits was not integrated. Instead, we looked here for  collisions of   
ellipses  which fulfil the conservation of mass and of linear momentum. These simulations were thus started with a 
large number of initial bodies (10 000) and continued until their final evolution.
The different runs resulted in different numbers of planets, which ranged from four to nine, but the averaged values 
fitted very well with the  analytical  results of Table~\ref{table a_n} \citep{Laskar2000}.

\section{AMD-classification of planetary systems}
\llabel{classification}

Here we  show how the AMD-stability can be used as a criterion to derive a classification of planetary systems.
In section \ref{C_c} we saw that in the secular approximation, the stability of a pair of planets is determined by the computation of
\begin{equation}
\beta = \frac{\cC}{C_c}=\frac{C}{\Lambda'C_c} \ .
\llabel{beta}
\end{equation}
We  call $\beta$ the AMD-stability coefficient. For pairs of planets, $\beta<1$ means that collisions are not possible. 
The pairs of planets are then  called AMD-stable.
We  naturally extend this definition to multiple planet systems. 
A system is AMD-stable if every adjacent pair is AMD-stable\footnote{This is equivalent to require that all pairs are AMD-stable.}.
We can also define an AMD-stability coefficient regarding  the collision with the star. \LEt{ with regard to }
We define $\beta_S$, the AMD-stability coefficient of the pair formed by the star and the innermost planet. 
For this pair, we have $\alpha=0$ and thus $C_c=1$. 
With this simplification $\beta_S=C/\Lambda$, where $\Lambda$ is the circular momentum of the innermost planet.

\subsection{Sample studied and methods of computation}

We have studied the AMD-stability of some systems referenced in the exoplanet.eu catalogue \citep{Schneider2011}. 
From the catalogue, we selected  \totnb systems that have measured masses, 
semi-major axes, and eccentricities for all their planets. Since the number of systems with known mutual inclinations is too small, we assumed the systems to be almost coplanar. 
This claim is supported by previous statistical studies that constrain the observed 
mutual inclinations distribution (\citealp{Fang2012,Lissauer2011a,Fabrycky2012}; and \citealp{Figueira2012}). 
For some systems where the uncertainties were not provided,
we consulted the original papers or the Exoplanet Orbit Database\footnote{http://exoplanets.org/} \citep{Wright2010}.

\begin{figure}
        \begin{center}
                \includegraphics[width=\jxllinewidth]{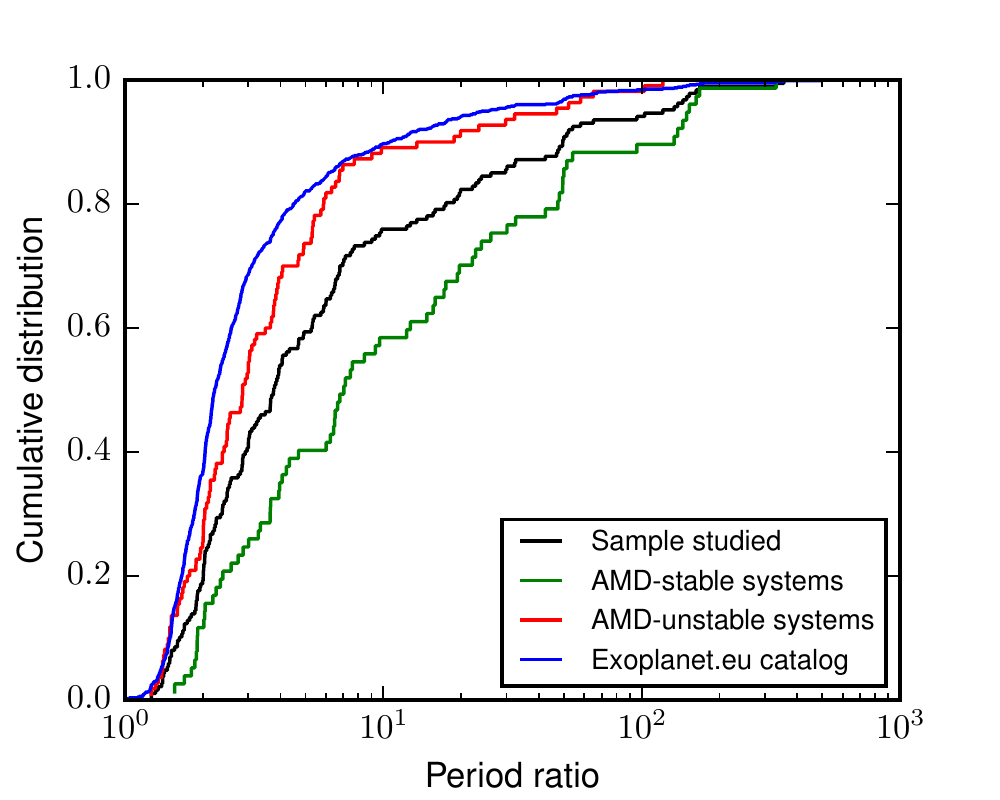}
                \caption{Cumulative distributions of the period ratios of adjacent planets in the sample studied here,
                \LEt{ yes? or: the adjacent planets' period ratios } of the AMD-stable systems (both weak and strong), 
                AMD-unstable systems, and for all the systems referenced in the exoplanet.eu database. }
                \label{CDFper}
        \end{center}
\end{figure}
We compare the cumulative distribution of the adjacent planets' period ratios in our sample
and that of all the multiplanetary systems in the database exoplanet.eu (see Fig.~\ref{CDFper}). 
The sample is biased toward higher period ratios.
Indeed, most of the multiplanetary systems in the database come from the Kepler data.
Since these systems are mostly tightly packed ones, their period ratios are rather small.
However, the majority of them do not have measured eccentricities and are consequently excluded from this study.
Our sample thus mostly contains   systems detected by radial velocities (RV) methods that have, on average, higher period ratios.

Since all the AMD computations are done with the relative quantities $\alpha$ and $\gamma$, we can use equivalent quantities that are  measured more precisely in observations than the masses and 
semi-major axis. We used the period ratios elevated to the power 3/2 instead of the semi-major axis ratios,
and  the minimum mass $m\sin(i)$ for RV system. This is not a problem for the computation of $\gamma$; if we assume that the systems are close to coplanarity, then
\be
\gamma=\frac{m}{m'}=\frac{m\sin(i)}{m'\sin(i)}\simeq\frac{m\sin(i)}{m'\sin(i')}\ .
\ee

\begin{figure}
        \begin{center}
                \includegraphics[width=\jxllinewidth]{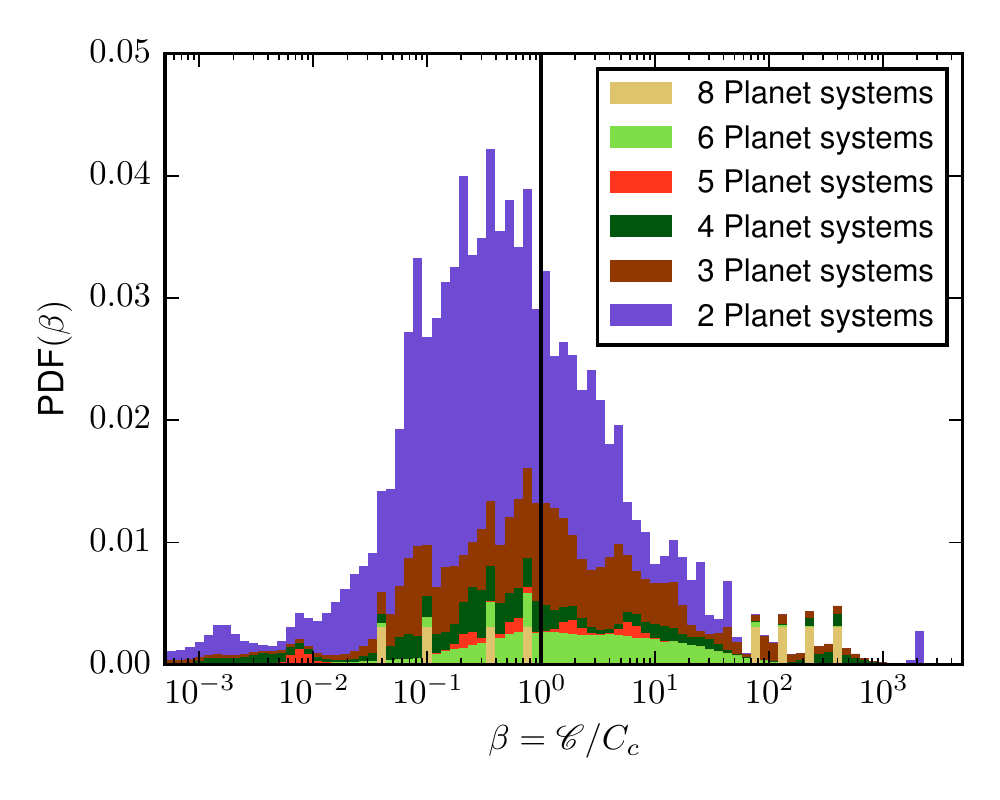}
                \caption{Probability distribution function of $\beta$ for the sample studied. 
                The systems are grouped by multiplicity. The vertical line $\beta=1$ marks the separation between 
                AMD-stable and AMD-unstable pairs.}
                \label{CDFbeta}
        \end{center}
\end{figure}

Even though we assume the systems to be coplanar, we want to take into account the contribution of mutual inclinations to the AMD.
Since we only have  access to the eccentricities, we define the coplanar AMD of a system $C_p$, as the AMD of the same system if it was coplanar. \LEt{ ambiguous here: 1) Since we only have access to the eccentricities, if it is coplanar we define the coplanar AMD of a system $C_p$ as the AMD of the same system. 2)
Since we only have access to the eccentricities, we define the coplanar AMD of a system $C_p$ as if the AMD of the same system were coplanar.
  }
We can also define 
\be
\beta_p = C_p/(\Lambda'C_c) \ ,
\llabel{copAMDcoeff}
\ee
which is  the coplanar AMD-stability coefficient. Motivated  by both theoretical arguments 
on chaotic diffusion in the secular dynamics \citep{Laskar1994,Laskar2008} and observed correlations in statistical distributions \citep{Xie2016}, 
we assume that the AMD contribution of mutual inclinations is equal to that of the eccentricities.
This  \ap{hypothesis is equivalent to assuming  the} average equipartition of the AMD \ap{among} the secular degrees of freedom for a chaotic system. The total AMD is thus twice the measured AMD, and in this study we  use  
\be \beta=2\beta_p \ .\ee
We can also define a coplanar AMD-stability coefficient associated with the star, and similarly we set $\beta_S=2\beta_{Sp}$.
We then compute the coefficients $\beta_S$ and $\beta$ for each pair and the associated uncertainty distributions.
We list the results of the analysis in Table~\ref{datarep}. In the considered 
dataset, \totsta systems are AMD-stable. The majority of the  highest multiplicity systems are  AMD-unstable. 
\tableD
In Figure~\ref{CDFbeta} we plot the probability distribution of $\beta$ for the considered systems.

\subsection{Propagation of uncertainties }

The uncertainties are propagated using Monte Carlo simulations of the distributions.
After determining the distributions from the input quantities ($m,a,e$),
we generate 10,000 values for each of these parameters.
We then compute the derived quantities ($\alpha,\gamma,C_c,\beta$\dots) in these 10,000 cases and the associated distributions.
 
For masses (or $m\sin(i)$ if no masses are provided) and  periods, we assume a Gaussian uncertainty centred on the value referenced in the database and with standard deviation, the half width of the confidence interval.
The distributions are truncated to 0.

For eccentricity distributions, the previous method does not provide satisfying results.
Most of the Gaussian distributions constructed with the mean value and confidence interval given in the catalogue make  negative eccentricities probable (in the case of almost circular planets with a large upper bound on the eccentricity).
One solution is to assume that the rectangular eccentricity coordinates $(e\cos \omega,e\sin \omega)$ are Gaussian.
Since the average value of $\omega$ has no importance in the computation of the eccentricity distribution, we assume it to be $0$.
Therefore, $e\sin\omega$ has 0 mean. 
We define the distribution of $\tilde e=e\cos \omega$ as a Gaussian distribution with the  mean value, the value referenced in the catalogue, and standard deviation, the half-width of the confidence interval.
If we assume $e\sin\omega$ has the same standard deviation as $e\cos \omega$, we have $e\sin\omega=\tilde e -\av{\tilde e}$.
The distribution of $e$ is then deduced from that of  $\tilde e$  using
\begin{equation}
        e=\sqrt{{\tilde e}^2+(\tilde e -\av{\tilde e})^2}\ .
\end{equation}

Using the Gaussian assumption means that some masses or periods can take values close to 0 with a small probability. \LEt{ a small probability=a few per cent. If you can give a specific number you can put the phrase back in }
This causes the distributions of $\alpha$ or $\gamma$ to diverge if it happens that $a'$ or $m'$ can take values close to 0.
To address this issue, a linear expansion around the mean value is used for the quotients, for example for $\alpha$,
\begin{equation}
\alpha=\frac{a}{a'}=\frac{a}{\av{a'}}\left(1-\frac{\Delta a'}{\av{a'}}\right) \ ,
\end{equation}
with $\Delta a'=a'-\av{a'}$.

\begin{figure*}[h!]
        \begin{center}
                \includegraphics[width=\jxllinewidthD]{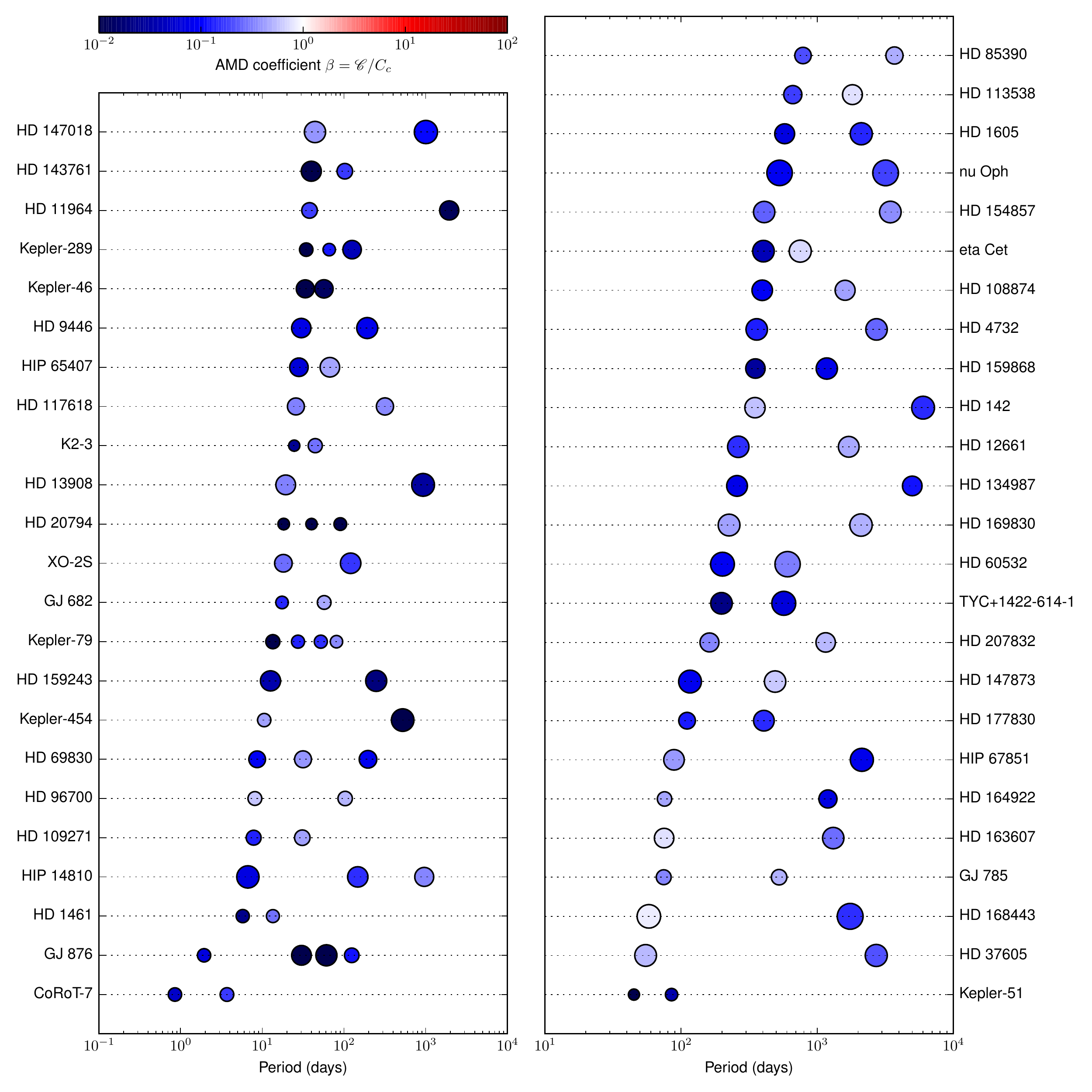}
                \caption{Architecture of the strong AMD-stable systems. Each planet is represented by a circle 
                whose size  is  proportional to $\log_{10}(m),$ where $m$ 
                is the mass of the planet. \LEt{ yes? also Figs. 6, 7 } The colour represents the 
                AMD-stability coefficient of the {inner} pair associated with the planet 
                (in particular, the first planet is represented with the AMD-stability coefficient associated with the star $\beta_S$). This means 
                that a red planet can collide with its inner neighbour}
                \label{strongstable}
        \end{center}
\end{figure*}

\subsection{AMD-stable systems}
\label{secsta}

As said above, we consider AMD-stable a system where collisions between planets are impossible because of the dynamics ruled by AMD ($\max \beta<1$).
In addition, if the AMD-stability coefficient of the star $\beta_S<1$ (resp. $\beta_S>1$), 
the system is defined as strong AMD-stable (resp.  weak AMD-stable).

\subsubsection{Strong AMD-stable systems}

The strong AMD-stable systems can be  considered  dynamically stable in the long term. In Figure~\ref{strongstable}, we plot  the architecture of 
the strong AMD-stable systems. 
If the system is out of the mean motion resonance (MMR) islands, the AMD will not increase and therefore no collision between planets or 
with the star can occur.
We can see in Figure~\ref{CDFper} that the AMD-stable systems have period ratios on average larger than those from the sample.

\subsubsection{Weak AMD-stable systems}

\begin{figure}[ht]
        \begin{center}
                \includegraphics[width=\jxllinewidthb]{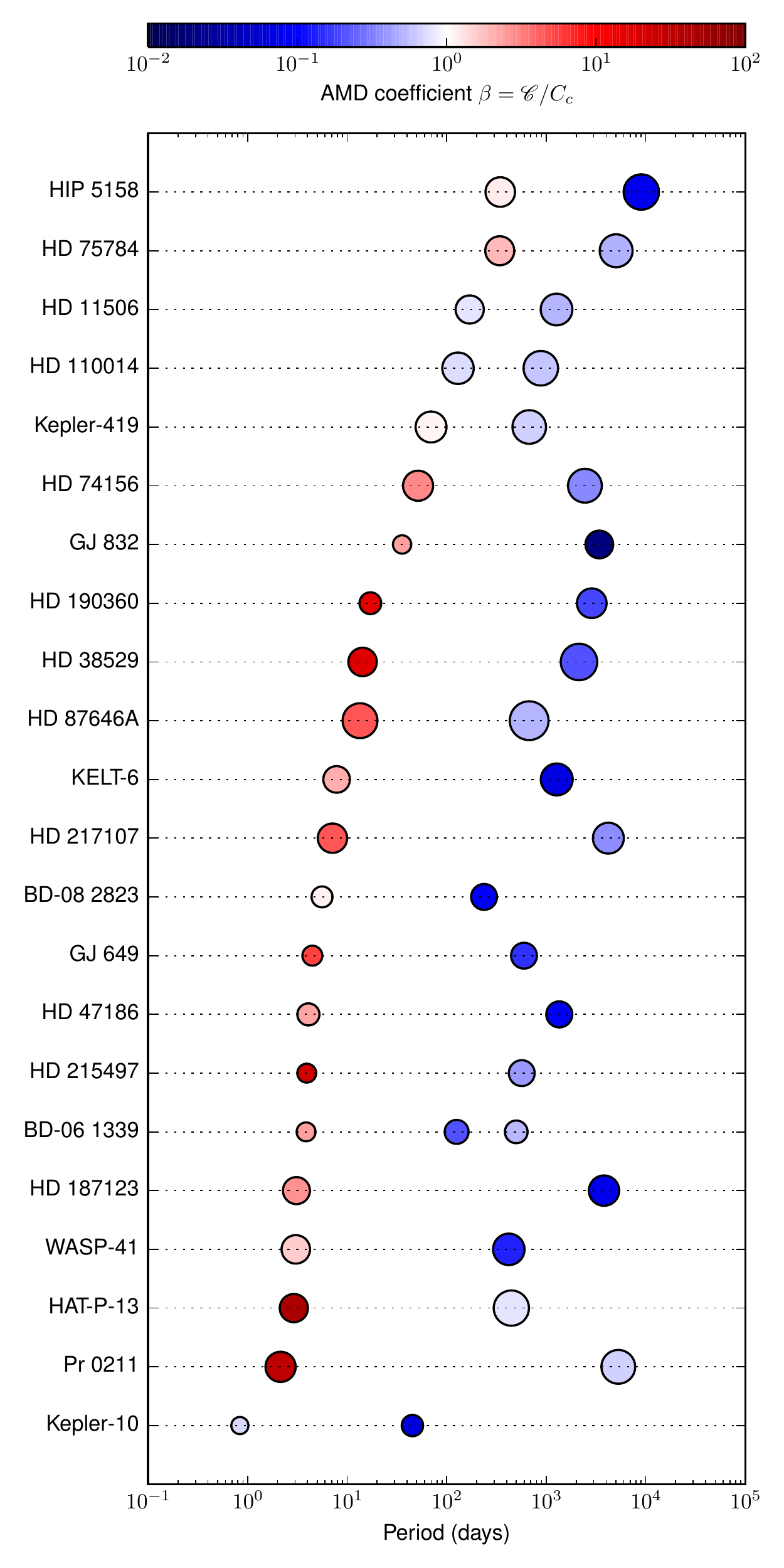}
                \caption{Architecture of the weak AMD-stable systems. Each planet is represented by a circle  
                whose size is proportional to $\log_{10}(m),$ where $m$ is 
                the mass of the planet. The colour represents the AMD-stability coefficient of the {inner} 
                pair associated with the planet (in 
                particular, the first planet is represented with the AMD-stability coefficient associated with the star $\beta_S$).}
                \label{weakstable}
        \end{center}
\end{figure}

As defined above, in a weak AMD-stable system, no planetary collisions can occur,
but the innermost planet can increase its eccentricity up to 1 and collide with the star. 
We separate these systems  from the strong AMD-stable ones because the system can still lose a planet
only by AMD exchange. However, the remaining system will not be affected by the destruction of the inner planet.
In Figure~\ref{weakstable}, we plot their architecture. 
In these systems, the inner planet is much closer to the star than the others.

\subsection{AMD-unstable systems}
\label{secuns}
\begin{figure*}[ht]
        \begin{center}
                \includegraphics[width=\jxllinewidthD]{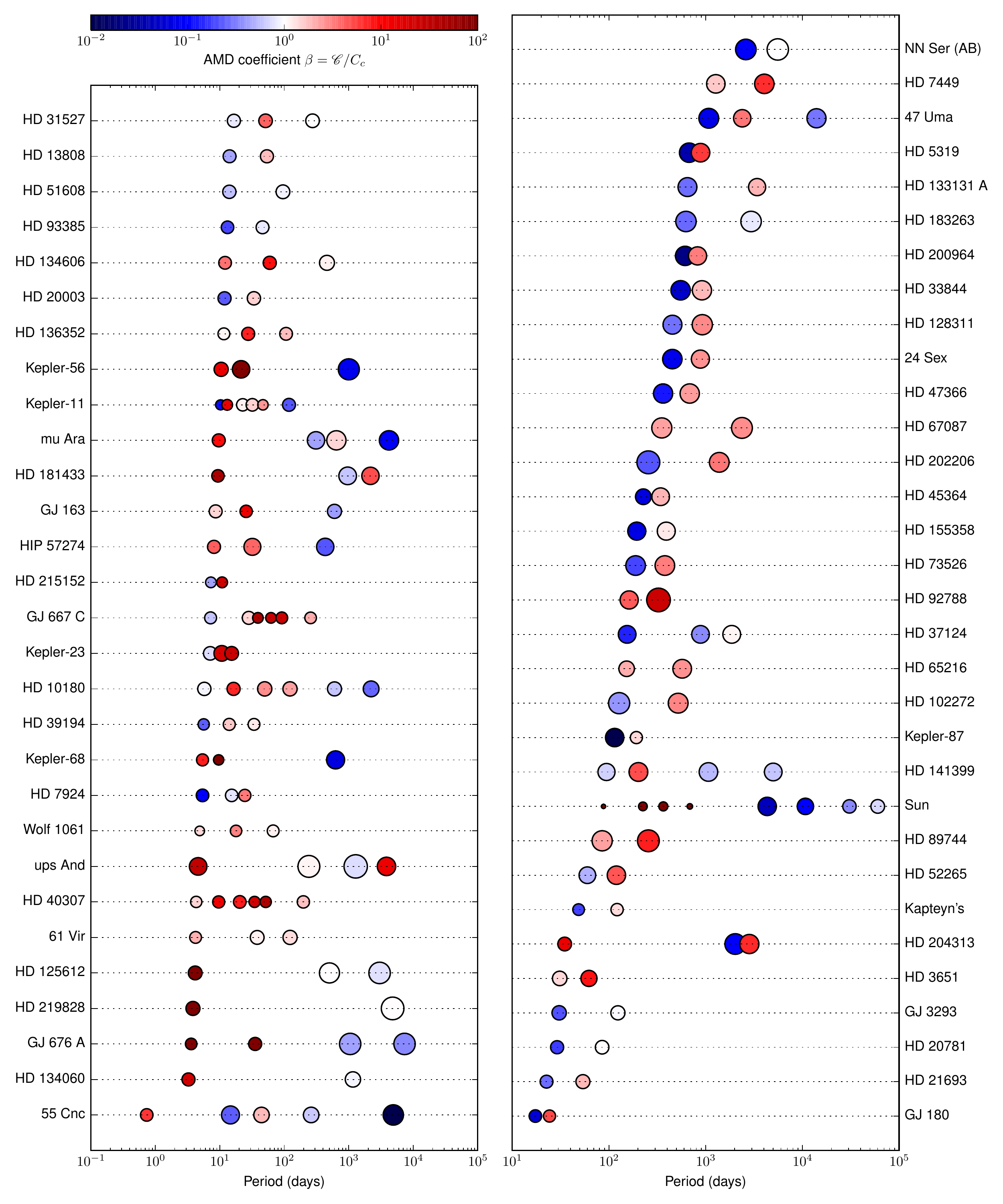}
                \caption{Architecture of the AMD-unstable systems. Each planet is represented by a circle  
                whose size is proportional to $\log_{10}(m),$ where 
                $m$ is the mass of the planet. The colour represents the AMD-stability coefficient of the {inner} 
                pair associated with the planet (in 
                particular, the first planet is represented with the AMD-stability coefficient associated with the star $\beta_S$).}
                \label{unstable}
        \end{center}
\end{figure*}

The AMD-unstable systems have at least one unstable planet pair,
but as we show in Figure~\ref{unstable} where we plot the architecture 
of these systems, this category is not homogeneous. 
It gathers high multiplicity systems where planets are too close to each other for the criterion to be valid, multiscale systems like the 
solar system where the inner system is AMD-unstable owing to its small mass in comparison to the outer part, systems experiencing mean 
motion resonances, etc. 
In all these cases, an in-depth dynamical study is necessary to determine the short- or long-term stability of these systems.
In the following sections, we  detail how the AMD-stability and AMD driven dynamics can help to understand these systems.

\subsubsection{Hierarchically AMD-stable systems (solar system-like)}

We first consider the solar system. Owing to the large amount of AMD created by the giant planets, the inner system is not AMD-stable 
\citep{Laskar1997}.
However,  long-term secular and direct integrations have shown that the inner system  has  a  probability of becoming 
unstable of only 1\%  over 5~Gyr \citep{Laskar2008,Laskar2009,Batygin2015}.
Indeed, the chaotic dynamics is mainly restricted to the inner system, while the outer system is mostly quasi-periodic. 
However, when AMD exchange occurs between the outer and inner systems, the inner system becomes highly unstable and can lose one or 
several planets.
\begin{figure}[htbp]
        \begin{center}
                \includegraphics[width=\jxllinewidth]{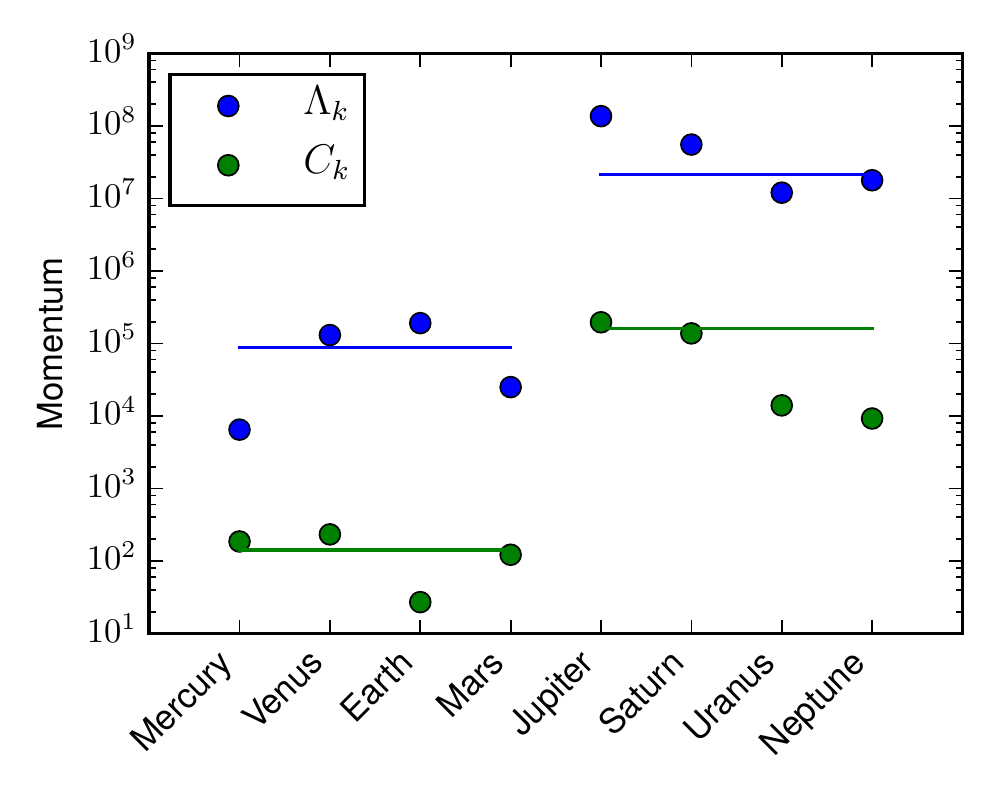}
                \caption{Circular momenta and AMD of the solar system planets. The horizontal 
                lines correspond to the respective mean $\Lambda_k$ 
                and $C_k$ for the inner and outer system.}
                \label{AMD SS}
        \end{center}
\end{figure}

In Figure~\ref{AMD SS} we plot  the AMD and the circular momenta of the solar system planets. 
We see that the inner system planets (resp. the outer planets) have comparable AMD values. \cite{Laskar2008} showed the absence of exchange 
between the two parts of the system.
Moreover, the spacing of the planets follows surprisingly well the distribution laws mentioned in section \ref{formation} if we consider the two parts of the system separately \citep{Laskar2000}.

\begin{figure}[htbp]
        \begin{center}
                \includegraphics[width=\jxllinewidth]{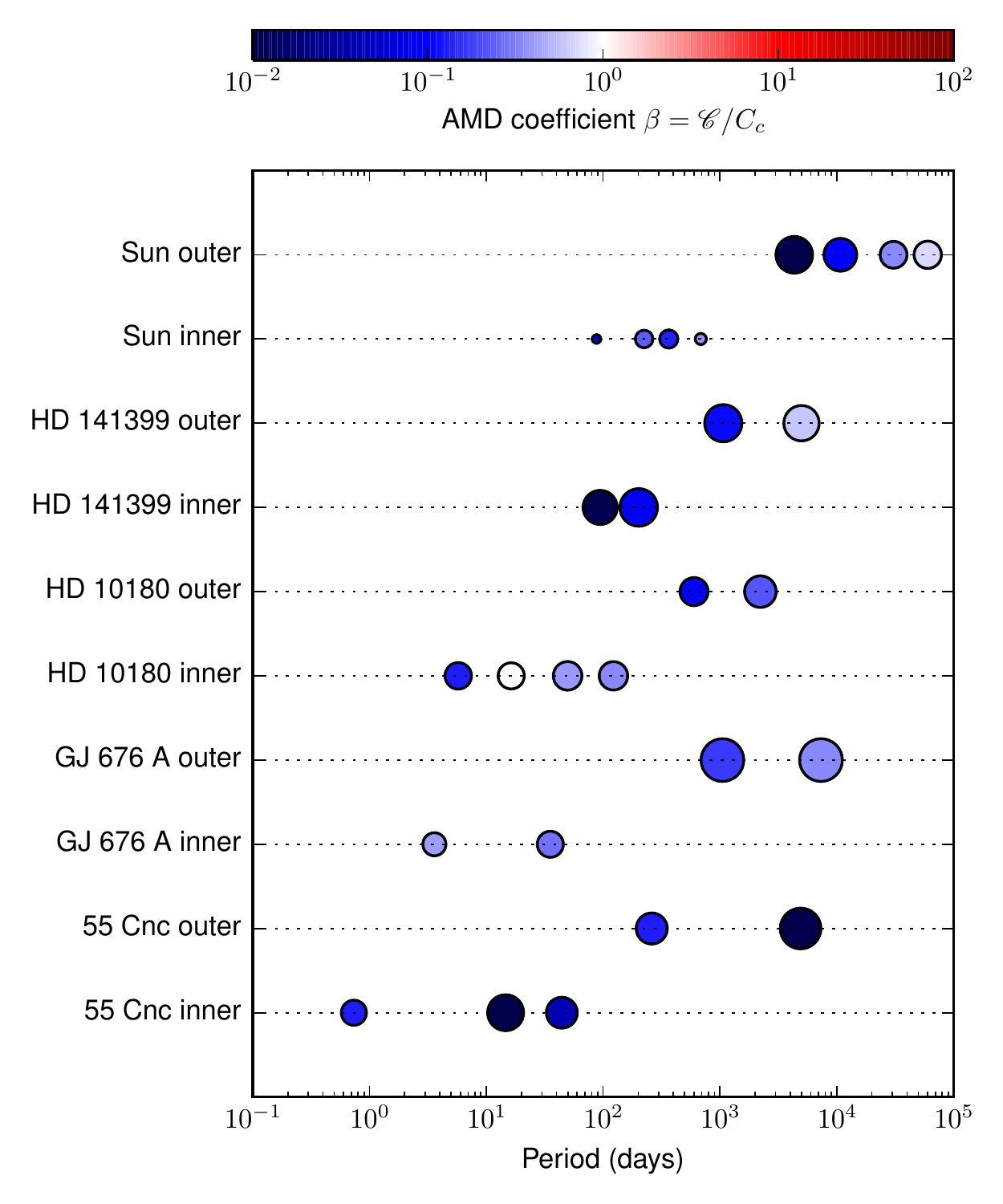}
                \caption{Examples of systems where the outer part is AMD-stable and the inner part becomes AMD-stable if considered alone. }
                \label{split}
        \end{center}
\end{figure}

We see that given an analysis of the secular dynamics, the tools developed above can still help to understand how an {a priori} 
unstable system can still exist in its current state.
The case of AMD-unstable systems with an AMD-stable part is not restricted to the solar system.
If we look for systems where the outer part is AMD-stable while the whole system is AMD-unstable, we find four similar systems in our 
sample as shown in Figure~\ref{split}.
We  call these systems hierarchic AMD-stable systems.

\subsubsection{Resonant systems}

As shown by the cumulative distributions plotted in  Figure~\ref{CDFper}, the AMD-unstable systems have period ratios that are lower than the 
AMD-stable systems. 
For example, in our sample 66\% of the adjacent pairs of the AMD-unstable systems  have a period ratio below 4, whereas this proportion is 
 only 33\% for the AMD-stable systems.
For period ratios close to small integer ratios, the MMR can rule a great part of the dynamics.
Particularly, if pairs of planets are trapped in a large resonance island,
the system could be dynamically stable even if it is not AMD-stable.

Individual dynamical studies are necessary in order to claim that a system is stable due to a MMR.
We note, however,  that the AMD-unstable systems considered here are statistically constituted of more resonant pairs than a typical 
sample in the catalogue. 

\subsubsection{Concentration around MMR}

\begin{figure}
        \begin{center}
                \includegraphics[width=\jxllinewidth]{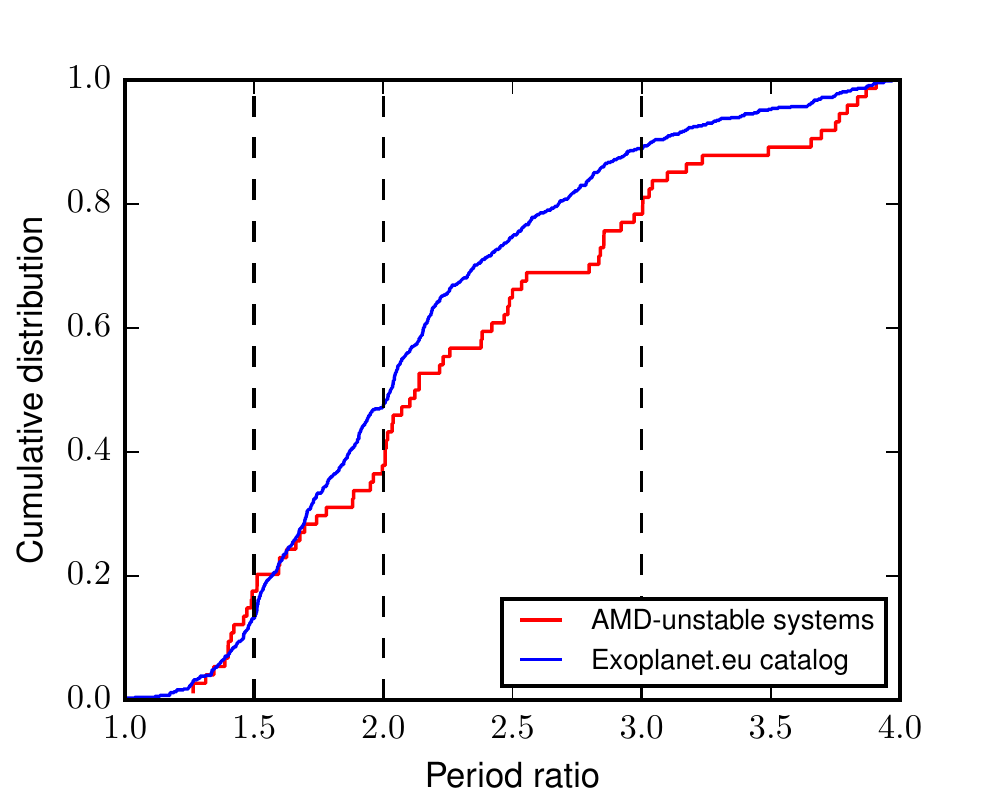}
                \caption{Cumulative period ratios of the AMD-unstable systems and of all systems in the catalogue.
                The vertical line represents low order MMR.}
        \label{CDFper4}
        \end{center}
\end{figure}

Here we want to test whether the unstable systems have been drawn randomly 
from the exoplanet.eu catalogue or if \ap{the} period ratios \ap{of the pairs of adjacent planets} are statistically closer to small integer ratios.
We denote $\H_0$, the hypothesis that the period ratios  of the unstable systems have been drawn randomly from the catalogue 
distribution.
We consider only the period ratios lower than 4 because the higher ones are not significant for a study of the MMR.
We plot in figure~\ref{CDFper4} the cumulative distribution of the period ratios of the AMD-unstable systems, as well as the cumulative 
distribution of the period ratios of all the systems of exoplanet.eu.
We \ap{call} $R_u$ the set of period ratios of the AMD-unstable planets.

We first test $\H_0$ via  a Kolmogorov-Smirnov test \citep{Lehmann2006} between the sample $R_u$ and the period ratios of the 
catalogue.
The test fails to reject $\H_0$ with a p-value of about 9\%.
Therefore, we cannot reject $\H_0$ on the global shape of the distribution of the AMD-unstable period ratios.

However, we want to determine if the AMD-unstable pairs are close to small integer ratios, which means studying the fine structure of the 
distribution.
We propose here another method for demonstrating this.

Let us denote the probability density of the catalogue period ratios $f$ and the associated cumulative distribution $\F$.
Let us consider an interval $I_x=(x,x+\Delta x)$, if we assume $\H_0$, the probability for a ratio $r$ to be in $I_x$ is 
\begin{eqnarray}
\Pr(r\in I_x|\H_0)&=&\int_{x}^{x+\Delta x}f(r')\d r'\nnb
                  &=&\F(x+\Delta x)-\F(x)=\Delta \F(x) \ .
\end{eqnarray}

Therefore, the probability that more than $k_0$ pairs out of $N$ pairs drawn randomly from the distribution $\F$ are in $I_x$ is
\begin{equation}
\Pr(x,k_0|\H_0)=\sum_{k=k_0}^{N}\binom{N}{k}(1-\Delta \F)^{N-k}\Delta \F^k.
                \label{Pxk0}
\end{equation}
This is the probability of a binomial trial. From now on, $N$ \ap{designates} the number of period ratios in $R_u$.

\begin{figure}
        \begin{center}
                \includegraphics[width=\jxllinewidth]{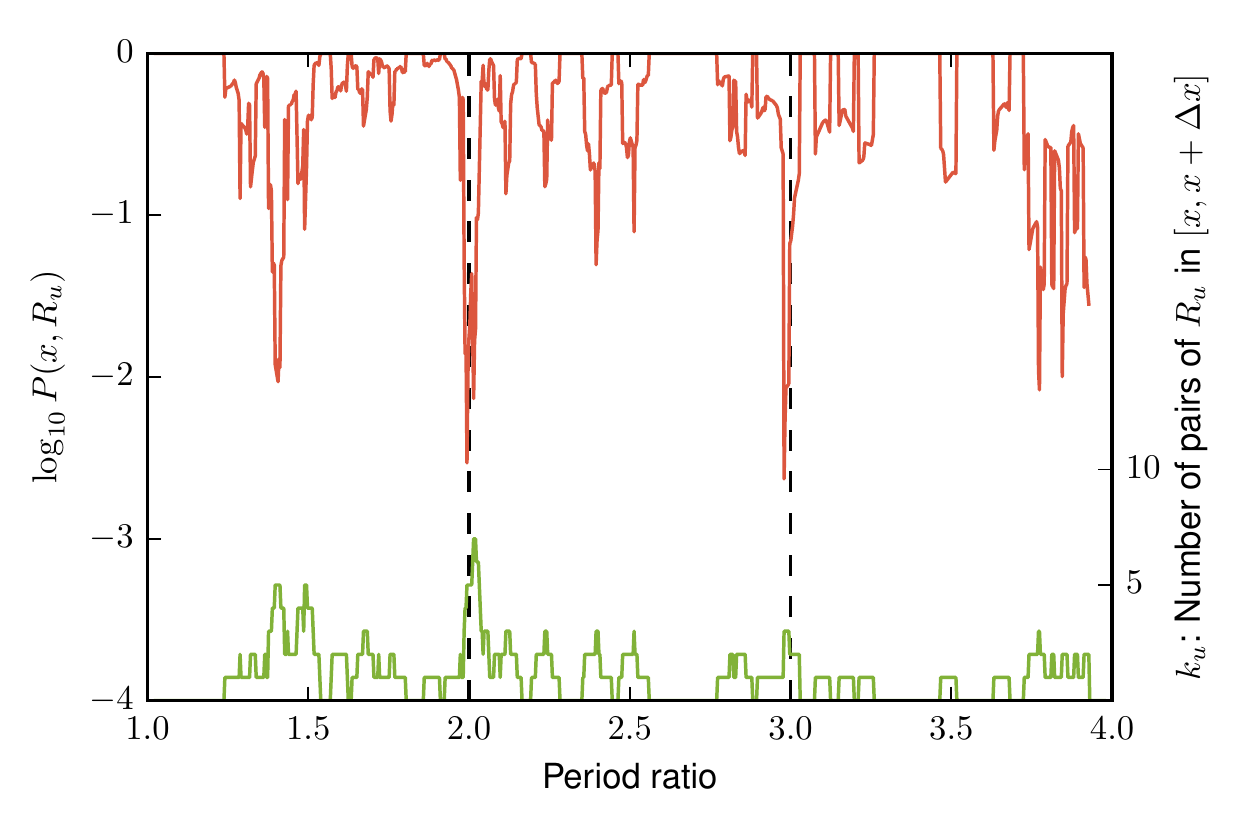}
                \caption{Red curve: Probability of observing more than $k_0$ pairs in the interval $I_x$ given a probability density $f$.
                Green curve: $k_0$ is the number of pairs in $I_x$ for the unstable systems $R_u$}
        \label{Pxk0fig}
        \end{center}
\end{figure}
For a given $\Delta x$, we can compute the probability $P(x,R_u)=\Pr(x,k_u(x))$, where $k_u(x)$ is the number of pairs from $R_u$ in $I_x$. 
This probability measures the likelihood of a concentration of  elements of $R_u$ around $x$, assuming they were drawn from $\F$.
We choose $\Delta x =0.05$ and plot the function $P(x,R_u)$  as well as $k_u$ in Figure~\ref{Pxk0fig}. 
We observe that the concentrations around 3 and 2 are very unlikely with a probability of $2.3\times10^{-3}$ and $2.9\times10^{-3}$, respectively.
Other peaks appear for $x=1.4$ and around \ap{3.8}.
However, $P(x,R_u)$ is the probability of seeing a concentration around $x$ precisely.
It is not the probability of observing a concentration somewhere between 1 and 4.

\begin{figure}
        \begin{center}
                \includegraphics[width=\jxllinewidth]{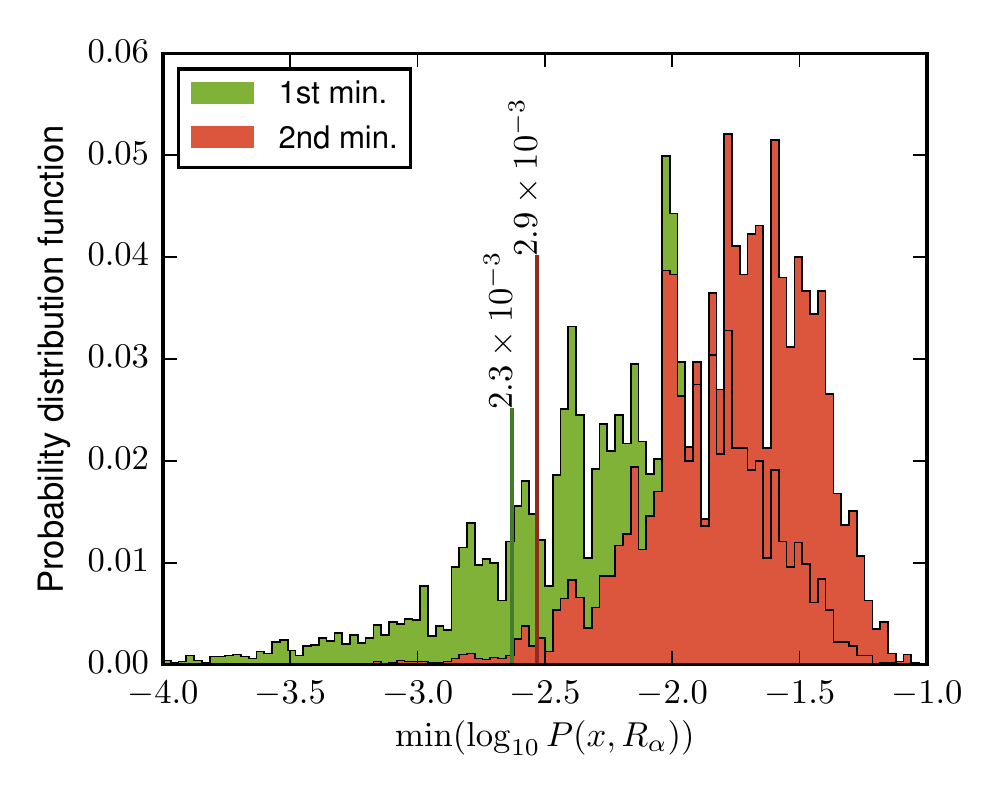}
                \caption{Probability density function of the two first minimum values of $P(x,R_\alpha)$ from 10,000 samples $R_\alpha$ drawn from the $\F$ distribution.}
                \label{pdfpvalue}
        \end{center}
\end{figure}

To demonstrate that the concentrations around 2 and 3 are meaningful, we compare this result to samples drawn randomly from the 
distribution $\F$. 
We create 10,000 datasets $R_\alpha$ by drawing $N$ points from $\F$ and compute $P(x,R_\alpha)$ for these datasets. 
Then, we record the two minimum values (at least distant by  more than $\Delta x$) \LEt{ minimum distance greater than $\Delta
x$ ? } and plot the distribution of these minima on 
Figure~\ref{pdfpvalue}. 
From these simulations, we see that on average 17.2\% of the samples have a minimum as small as $R_u$.
However, the presence of a secondary  peak as significant as the second one of $R_u$ has a probability of 1.3\%.
Moreover, the $R_u$ concentrations are clearly situated around low integer ratios, which would not be the case in general for a randomly 
generated sample.

We thus demonstrated here that the AMD-unstable systems period ratios are significantly concentrated around low integer ratios.
While we do not prove that the pairs of planets close to these ratios are actually in MMR, this result further motivates  investigations toward the behaviour of these pairs in a context of secular chaotic dynamics.

\section{Conclusion}

The angular momentum deficit (AMD) \citep{Laskar1997,Laskar2000} is a key parameter for the understanding 
of the outcome of the formation processes of planetary systems \citep[e.g.][]{Chambers2001,Morbidelli2012,Tremaine2015}.
We have shown here how AMD can be used to derive a well-defined stability criterion:  the AMD-stability. 
The AMD-stability of a system can be checked by the computation of  the critical AMD \ap{$C_c$ (eq. \ref{colli})} and AMD-stability coefficients $\beta_i$ 
that depend only on the eccentricities and ratios of semi-major axes and masses (Eqs. \ap{\ref{amd},} \ref{colli}, \ref{critic}, \ref{beta}).
This criterion  thus does not depend on the degeneracy of the masses coming from radial velocity measures. Moreover, the uncertainty 
on the relative  inclinations  can be compensated by assuming equipartition of the AMD between eccentricities and inclinations. 

AMD-stability will ensure that in the absence of mean motion resonances, the system is long-term stable. A  rapid estimate of the stability 
of a system can thus be obtain by a short-term integration and the simple computation of the AMD-stability coefficients. 

We have also proposed here a classification of  the planetary systems based on AMD-stability (Section \ref{classification}). 
The  strong AMD-stable systems  are the systems where no planetary collisions are  possible, and no collisions of the inner planet  with 
its central star, while the weak AMD-stable systems allow for the collision of the inner planet  with the central star. 
The AMD-unstable systems are the systems for which the AMD-coefficient does not prevent the possibility of collisions. The solar system 
is AMD-unstable, but it belongs to the subcategory  of hierarchical AMD-stable systems that  are AMD-unstable but 
 become AMD-stable when they are split  into two parts (giant planets and terrestrial planets for the solar system) \citep{Laskar2000}. 
Out of the 131 studied systems from exoplanet.eu, we find 48 strong  AMD-stable,  22 weak AMD-stable, and 61 AMD-unstable systems, including 5 hierarchical AMD-stable systems. 

As for the solar system, the AMD-unstable systems are not necessarily unstable, but  determining their stability requires some 
further dynamical analysis. Several mechanisms can stabilise AMD-unstable systems.
The absence of secular chaotic interactions between parts of the systems, like in the solar system case, 
or the  presence of mean motion resonances, protecting pairs of planets from collision.
In this case, the AMD-stability classification is still useful in order to select the systems that require this more in-depth 
dynamical analysis. 
It should also be noted that the discovery of additional planets in a system will require a revision of the computation of the  AMD-stability of the system. This additional planet will always increase the total AMD, and thus  the maximum AMD-coefficient  of the system, decreasing its AMD-stability unless it is split into two subsystems. 

In the present work, we have not taken into account mean motion resonances (MMR) and  the chaotic behaviour resulting from their overlap. 
We aim to take these MMR  into consideration in  a forthcoming work. Indeed,  
the criterion \LEt{ or criteria? } regarding MMR developed by \cite{Wisdom1980} or more recently \cite{Deck2013} may help to 
improve our stability criterion by considering the MMR chaotic zone as a limit for stability instead of the 
limit considered here that is  given by the collisions of the orbits (Eqs. \ref{colli}). 
The drawback  will then most probably be giving up \LEt{ ignoring? renounce is different and not appropriate here }  the  rigorous results that we have established   in section  
\ref{sec.AMD stab},
and  allowing for additional empirical studies.\LEt{  or studies that are more empirical ?   } The present work will in any case remain the backbone of these 
further studies. 
Note: The AMD-stability coefficients of selected planetary systems will be made available on the exoplanet.eu database.

%=========%=========%=========%=========%=========%=========%=========%=
%  BIBLIOGRAPHIE
%=========%=========%=========%=========%=========%=========%=========%=
\bibliographystyle{aa}  % A&A bibliography style file (aa.bst)
\bibliography{AMDAA} % references .bib file

\appendix

\renewcommand\appendixname{Appendix}

\section{Intersection of planar orbits}

In this section, we present an efficient algorithm for the computation of the
intersection of two elliptical orbits in the plane, following 
\citep{Albouy2002}.
Let us consider an elliptical orbit defined by $(\mu, \r, \dr)$ and let 
$\G = \r\wedge \dr$ be the angular momentum \ap{per} unit of mass.
The Laplace vector
\be
\P = \Frac{\dr\wedge \G}{\mu}- \Frac{\r}{r}
\ee
is an integral of the motion with coordinates $( e\cos \omega,e\sin \omega)$. One has
\be
\P\cdot \r = \frac{G^2}{\mu} - r = p-r
\llabel{eq.19}
,\ee
where $p= a(1-e^2)$ is the parameter of the ellipse. Let $\r =(x,y)$ in the plane. We
can consider the ellipse in three-dimensional space $(x,y,r)$ as the intersection  of
the cone 
\be
r^2 = x^2 + y^2
\ee
with the plane defined by Eq. (\ref{eq.19}), that is  
\be
x (e\cos \omega) + y (e\sin \omega) +r =p
\ee
If we consider now two orbits $\cO_1$ and $\cO_2$. Their intersection is easily obtained
as the intersection of the line of intersection of the two planes 
\be
\EQM{
x (e_1\cos \omega_1) + y (e_1\sin \omega_1) +r =p_1 \crm
x (e_2\cos \omega_2) + y (e_2\sin \omega_2) +r =p_2 \crm 
}
\ee
with the cone of equation $r^2=x^2 +y^2$. Depending on the initial conditions, if
$\cO_1$ and $\cO_2 $ are distinct,  we will  get either 0, 1, or 2 solutions.

\begin{figure}[h]
\begin{center}
\includegraphics*[width=6cm]{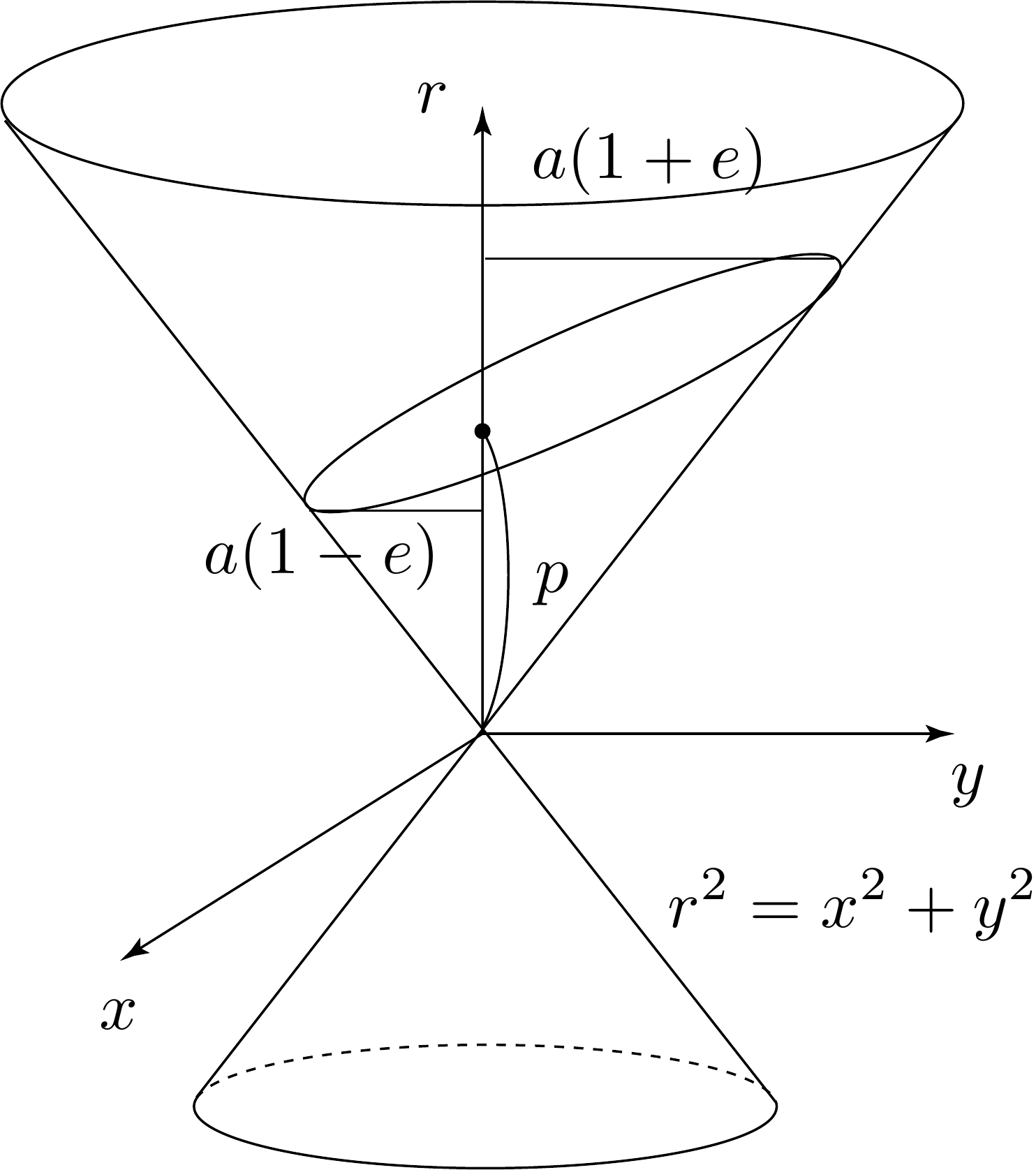}
\end{center}
\caption{Elliptical orbit, as the intersection of the cone $r^2 = x^2 + y^2$ 
and the  plane $\P\cdot \r +r = p$ .}
\label{fig1}
\end{figure}

\section{ AMD in the averaged equations}
\label{annex-average}
In this section we show that the AMD conserves the same form in the averaged planetary
Hamiltonian at all orders. More generally, this is true for any integral of $H$ 
which does not depend on the longitude $\l_k$.
Let 
\be
H = H_0(\L) + \ve H_1(\L,\l,J,\phi)
\ee
be a perturbed Hamiltonian system, and let $K(\L,J,\phi)$ be a first integral of $H(\L,\l,J,\phi)$
 (such that $\{K,H\}=0$, where $\{\cdot,\cdot\}$ is the usual Poisson bracket), and independent of $\l$.
In addition,  let 
\be
H' = \e^{L_S} H
\ee
be a formal averaging of $H$ with respect to $\l$. If $S(\L,\l,J,\phi)$ is a 
generator  defined as below (Eqs.~\ref{eq.hia},~\ref{eq.hib},~\ref{eq.hic}),  such that $H'$ is independent of $\l$.
Then, $K$ is an integral of $H'$, {i.e.}
\be
\{K,H'\}=0
\ee

The generator $S = \ve S_1+\ve^2 S_2+\ve^3 S_3 +\cdots $ is obtained formally through the identification 
order by order
\be
\EQM{
H'_0 = H_0 &\crm
H'_1 = H_1 +  & \{H_0,S_1\} \crm
H'_2 =      & \{H_0,S_2\}  + \Frac{1}{2} \{\{ H_0,S_1\},S_1\} +\{H_1,S_1\}\crm
\dots &
}
\llabel{eq.hia}
\ee

For any function $G(\L,\l,J,\phi)$, let 
\be{\langle G\rangle_{\bm\l} =\frac{1}{(2\pi)^m}}\int  G \d^m\bm{\lambda} \ee 
be the average of 
$G$ over all the angles $\l_k$. For each $n \geq 1$, the $S_n$ is obtained 
through the resolution of an  equation of the form 
\be
H'_n = \{H_0,S_n\} + {\cal R}_n
\llabel{eq.hib}
,\ee
where ${\cal R}_n $ belongs to ${\cal L}(H_0,H_1,S_1,\dots,S_{n-1})$, the  Lie algebra 
generated by $(H_0,H_1,S_1,\dots,S_{n-1})$. $H'_n$ will be  the averaged part of ${\cal R}_n$,
$\langle {\cal R}_n\rangle_\l $, and 
$S_n$ is  obtained by solving the homological equation
\be
\{H_0,S_n\} = \langle {\cal R}_n\rangle_{\bm\l}-{\cal R}_n \ .
\llabel{eq.hic}
\ee

We  show by recurrence that $\{K,S_n\} = 0$ for all $n \ge 1$.
First, we note that as $\{K,H_0\}=0$, we also have  $\{K,H_1\}=0$.  As $K$ is independent of 
$\l_k$, we also have  for all $G$
\be
\langle \{K,G\}\rangle_\l =\{K,\langle G\rangle_\l\}
\llabel{eq.aver}
\ee
This can be seen by formal expansion of $G$ in a Fourier series $G = \sum g_k \e^{i\langle k,\l\rangle}$. We have thus $\langle G\rangle_\l = g_0$. Let us  assume now that  $\{ K,S_k\} =0$  for all $k\leq n$. As
${\cal R}_{n+1} \in {\cal L}(H_0,H_1,S_1,\dots,S_{n})$, we also have  $\{K,{\cal R}_{n+1}\} =0$,
and from (\ref{eq.aver}) $\{ K, \langle {\cal R}_{n+1}\rangle_\l\}=0$ and thus 
\be
\{K,\{H_0,S_{n+1}\}\} = 0
\ee
Using the  Jacobi identity, as $\{F,H_0\}=0$, we have   
\be
\{H_0,\{K,S_{n+1}\}\} = 0
\llabel{eq.h0}
\ee
The solution  of the homological equation \ap{(\ref{eq.hic})} is  unique up to a term independent 
of $\l$. But as ${\langle \{K,S_{n+1}\}\rangle_\l =0}$, then the only possible solution for (\ref{eq.h0})
is 
\be
\{K,S_{n+1}\} =0 \ .
\ee
In the same way, as $H'_1 = \langle H_1\rangle_\l$, we have  $\{K,H_1'\} = \langle \{K,H_1\}\rangle_\l = 0 $. Thus
$\{K,\{H_0,S_1\}\} = 0$, by the Jacobi identity,  $\{H_0,\{K,S_1\}\} = 0$, and 
as previously, $\{K,S_1\} = 0$. Our recurrence is thus complete  and 
 it follows immediately that $\{K,H'\} =0 $.

\section{ Special values of $C_c(\a,\ga)$}
\label{Annex C_c}
This appendix provides the detailed computations and proofs of the results of section~\ref{C_c}
\subsection{Asymptotic value of $C_c(\a,\ga)$ for $ \ga \rightarrow 0$}
We have shown that $e_c(\a,\ga)$ is a differentiable function of $\ga$, which is monotonic 
(\ref{amds1.eq1}) and bounded ($e_c(a,\ga)\in [0,1]$). The limit 
$
e_{c}(\a,0) = \Lim{\ga \rightarrow 0} e_c(\a,\ga)
$
exists for all $\a\in ]0,1]$.
If $e_c(\a,\ga)$ is a solution of equation(\ref{colli}), it will also be 
a solution of the following cubic equation (in $e$), which is directly obtained from (\ref{colli}) 
by squaring, multiplication, and simplification by  $\a(1+e):$
\be
\EQM{
K(e,\a,\ga) = \a(\ga^2-\a)e^3 &-(2-\a)(\ga^2-\a) e^2 \crm
&-(1-\a^2)e +(1-\a)^2 = 0
}
\llabel{colli2}
\ee
As $e_c(\a,\ga)$, is a continuous function of $\ga$, when  $\ga \rightarrow 0$ 
the limits $e_{0c}(\a) $ 
will satisfy the limit equation 
\be
K_0(e,\a) =  (1-\a-\a e)^2(1-e ) = 0
\llabel{eqlim}
\ee
with   solutions $e_0=1/\a-1$ and $e_1=1$.
Depending on $\a$, several cases are treated:

\paragraph{$\bm{\a < 1/2}$ :} We have then  $e_0 > 1$, and the only possibility for $e_{c}(\a,0)$ is $e_1=1$; \LEt{ have I interpreted the punctuation correctly? }
as it is the only root of (\ref{eqlim}) which belongs to $[0,1],$ we have thus 
\be
\lim_{\ga
\rightarrow 0} e_c(\a,\ga) = 1
\ee
We have then 
\begin{eqnarray}
\lim_{\ga
        \rightarrow 0} e_c'(\a,\ga) &=& 1-2\a \ ; \nnb
\lim_{\ga
        \rightarrow 0} C_c(\a,\ga) &=& 1-2\sqrt{\a(1-\a)} \ ; 
\end{eqnarray}

In order to study the behaviour of $e_c(\a,\ga)$ in the vicinity of $\ga =0$, 
we can differentiate   $K(e_c(\a,\ga),\ga) =0$ twice, which gives
\be
\Der{e_c}{\ga}(\a,0) = 0 \, ; \qquad \Frac{d^2e_c}{d\ga^2}(\a,0) = \Frac{4(\a-1)}{(1-2\a)^2} <0 \ ,
\ee
thus $e_c(\a,\ga)$  decreases with respect to $\ga$ at $\ga=0$. 

The second order development of $C_c$ gives

\be
C_c(\a,\ga)=1-2\sqrt{\a(1-a)}+\ga\sqrt{\a}-\ga^2\sqrt{\frac{1}{\a}-1} +\O(\ga^3).
\ee

\paragraph{$\bm{\a > 1/2}$ :} In this case,  $e_0 < 1$. As $e_c(\a,\ga) \in ]0,e_0[$, we have 
$e_{c}(\a,0) \in  [0,e_0]$, which gives as the unique possibility 
\be
\lim_{\ga \rightarrow 0} e_c(\a,\ga) = e_0
\ee
with
\be
\lim_{\ga
\rightarrow 0} e_c'(\a,\ga) = 0 \ ; \qquad 
\lim_{\ga
\rightarrow 0} C_c(\a,\ga) = 0 \ . 
\ee

By setting $\ga=0$ in  (\ref{amds1.eq1}), we also obtain  
\be
\Der{e_c}{\ga}(\a,0) = -\Frac{e_0}{\a^{3/2}\sqrt{1-e_0^2}} < 0\ .
\ee

The development of $C_c$ gives

\be
C_c(\a,\ga)=\ga\left(\sqrt{\a}-\sqrt{2-\frac{1}{\a}}\right)-\Frac{\ga^2}{2\a-1}\Frac{(1-\a)^2}{\a} +\O(\ga^3).
\ee

\paragraph{ $\bm{\a = 1/2}$ :} In this case,  $e_0 = 1$, and the only solution is 
\be
\lim_{\ga
\rightarrow 0} e_c(\a,\ga) = 1
\ee
and
\be
\lim_{\ga
\rightarrow 0} e_c'(\a,\ga) = 0 \ ; \qquad 
\lim_{\ga
\rightarrow 0} C_c(\a,\ga) = 0 \ . 
\ee
Moreover, equation (\ref{colli2}) becomes
$  
\ga^2 2 e^2(3-e) = (1-e)^3
$.
We obtain thus the asymptotic value for  $e_c(1/2,\ga)$ when $\ga \rightarrow 0 $ as
\be
1-e_c(\frac{1}{2},\ga) \,\,{\sim}\,\, (4\ga)^{2/3}.
\ee

For $\a=1/2$, the development of $C_c$ in $\gamma$ contains non-polynomial terms in $\ga$ giving

\be
C_c(1/2,\ga)= \frac{\ga}{\sqrt{2}}-2^{-1/3}\ga^{4/3}-2^{-4/3}\ga^2 +\O(\ga^{8/3})
\ee

\subsection{Asymptotic value of $C_c(\a,\ga)$ for $ \ga \rightarrow +\infty$}
This case is more simple. If $e_c(\a,\ga)$ is a solution of Eq.~(\ref{colli}), then it will also be
a solution of eq. (\ref{colli2}), and thus of 
\be
\Frac{K(e,\a,\ga)}{\ga^2} =  0
\llabel{colli3}
\ee
As $e_c(\a,\ga)$ is monotonic and bounded, it has a limit when $\ga \rightarrow +\infty$, which
will verify the limit equation (\ref{colli3}), when  $\ga \rightarrow +\infty$, that is 
\be
K_\infty(\a,e) = e^2(2-\a-\a e) = 0
\llabel{colli4}
\ee
As $0<\a<1$, the only solution is $ e= 0$, and thus 
\be
\lim_{\ga 
\rightarrow +\infty} e_c(\a,\ga) = 0
\ee
and
\be
\lim_{\ga
\rightarrow +\infty} e_c'(\a,\ga) = 1-\a \ ; \quad 
\lim_{\ga
\rightarrow +\infty} C_c(\a,\ga) = 1-\sqrt{\a(2-\a)} \  
\llabel{eq.cinf}
\ee
and more precisely
\be
\EQM{
C(\a,\ga) = 1 &- \sqrt{\a(2-\a)} -\Frac{1}{2\ga}\Frac{\sqrt{\a}(1-\a)^2}{2-\a} \crm
&+\Frac{1}{\ga^2} \Frac{\sqrt{\a(2-\a)}(1-\a)^2}{2(2-\a)^3} + \O\left(\Frac{1}{\ga^3}\right)
}
\ee

\subsection{Study of $C_c(\a,\ga)$ for $\ga=1$ and $\ga = \sqrt{\a}$.}
For  $\ga=1$ or $\ga = \sqrt{\a}$, we can also obtain simple expressions 
for $C_c(\a,\ga)$. Indeed,
If $\ga = 1$, $K(e,\a,\ga)$ factorises in 
$(1-\a)(1+e)(\a e^2 -2e +1-\a) $ and we have a single solution for $e_c$ in the interval $[0,1]$, 
\be
\EQM{
e_c(\a,1) &= \Frac{1- \sqrt{1-\a+\a^2}}{\a} \ ; \crm
e_c'(\a,1) &= \sqrt{1-\a+\a^2} -\a \ ;
}
\ee
and 
\be
\EQM{
C_c(\a,1) = &\sqrt{\a} - \Frac{\sqrt{\a -2+ 2\sqrt{1-\a+\a^2}}}{\sqrt{\a}} \crm
&+ 1 -\sqrt{\a}\sqrt{1-2\a+2\sqrt{1-\a+\a^2}}
}
\llabel{eq.c1}\ee

For $\ga=\sqrt{\a}$, the cubic equation (\ref{colli2}) reduces to
\be
K(\a,\sqrt{\a}) = -(1-\a^2)e +(1-\a)^2 =0
\ee
with the single solution 
\be
e_c(\a,\sqrt{\a}) = e_c'(\a,\sqrt{\a})) = \Frac{1-\a}{1+\a}
\ee
and 
\be
C_c(\a,\sqrt{\a}) = (1-\sqrt{\a})^2
\llabel{eq.csa}
\ee

\subsection{$C_c(\a,\ga)$ for $ \a \longrightarrow 1$}

Let us denote $\eta = 1-\a$. The equation~\ref{colli2} can be developed in $\eta$,

\begin{eqnarray}
K(e,\a,\ga) &=&  (\ga^2-1)e^2(e-1) +\nnb
            &=&\eta e\left((\ga^2-1)(e-e^2)-2\right) +\nnb
            &=& \eta^2 (1-e-e^2-e^3) =0.
\label{colli5}
\end{eqnarray}

The zeroth and first orders of equation~\ref{colli5} imply that $e_c$ must go to zero; \LEt{ tend to zero?}
moreover, it scales with $\eta$. We write $e_c(\eta,\ga)=\kappa(\ga)\eta +\o(\eta)$.
We inject this expression in~\ref{colli5} and keep the second order in $\eta$

\begin{equation}
     (\ga^2-1) \kappa^2+2\kappa -1=0.
\end{equation}

We keep the solution that is positive and continuous in $\ga$ and we have
\begin{equation}
        e_c(\eta,\gamma)=\Frac{\eta}{\gamma+1} +\o(\eta).
\end{equation}

If we now compute $C_c$ developed for $\a \longrightarrow 1$ we have

\begin{eqnarray}
        C_c(1-\eta,\ga)&=&k(\gamma)\eta^2+\o(\eta^2)
                      \nnb&=&\Frac{1}{2}\Frac{\gamma}{\gamma+1}\eta^2 +\o(\eta^2)
\end{eqnarray}

%%%%%%%%%%%%%%  encadre

\end{document}